\documentclass[aps,prb,reprint,superscriptaddress]{revtex4-2}
\usepackage{graphicx}
\usepackage{hyperref}
\usepackage{xcolor}
\usepackage{amsmath}
\usepackage{amsfonts}
\usepackage{amssymb}
\usepackage{varioref}
\usepackage{float}
\usepackage{subfigure}
\usepackage{epstopdf}
\usepackage{braket}
\usepackage{physics}
\usepackage{datetime2}
\usepackage{subcaption}
\usepackage{float}
\usepackage{lipsum, babel}
\usepackage[toc,page]
{appendix}

\def\ii{\mathbf i}

\begin{document}
    \textheight=23.8cm

\title{
Quantization and quantum oscillations of the sublattice charge order in Dirac insulators}

\newcommand{\yerphiadd}{ A.I. Alikhanyan National Science Laboratory, Yerevan Physics Institute, 2 Alikhanyan brothers str, Yerevan 0036, Armenia}
\newcommand{\Nanjing}{ National Laboratory of Solid State Microstructures and Department of Physics, Nanjing University, Nanjing 210093, China}
\newcommand{\umassadd}{Department of Physics, University of Massachusetts, 710 North Pleasant Street, Amherst, MA 01003-9337, USA}

\author{Arindam Tarafdar}
    \email{atarafdar@umass.edu}
    \affiliation{\umassadd}

\author{Tigran A. Sedrakyan}
    \email{tsedrakyan@umass.edu}
    \affiliation{\umassadd}
    \affiliation{\yerphiadd}

\date{\today}

\begin{abstract}
We report the quantization, quantum oscillations, and singular behavior of sublattice symmetry-breaking sublattice charge order (SCO) in two-dimensional Dirac insulators at charge neutrality under perpendicular magnetic fields $B$. SCO is induced by staggered sublattice potentials, such as those originating from substrates, strains, hydrogenation, and chemical doping. 
In small non-quantizing magnetic fields that result in less than a flux quantum threading the system, and small sublattice symmetry breaking potentials, SCO exhibits perturbative singular magnetic field dependence, $\sim |B|$, originating from hopping between neighboring sites of the same sublattice. 
At intermediate magnetic fields, when the cyclotron gap between the zeroth Landau level and the first Landau level, $\omega_c$, is smaller than the sublattice potential, $\omega_c\lesssim \Delta$, SCO shows {\it universally} quantized plateaus owing to discrete Landau-level degeneracy. As the magnetic flux increases by one flux quantum, one electron (per spin) is transferred from the sublattice with a higher chemical potential to the sublattice with a lower chemical potential. 
One electron transfer between sublattices per flux quantum results from the sublattice polarization of the zeroth Landau level in gapped Dirac materials, realizing the topological Thouless pump effect.
At stronger magnetic fields, $\omega_c\gtrsim \ \Delta$, corresponding to integer quantum Hall regimes, SCO displays singularities based on the physics of quantum magneto-oscillations. Our findings suggest new ways to experimentally detect the presence of the energy gap in Dirac materials, irrespective of the gap size. 
\end{abstract}

\maketitle

\section{Introduction}

The influence of external magnetic fields on electronic materials has been a major topic in condensed matter physics for many decades. These studies have resulted in important discoveries of both theoretical and experimental importance, especially the quantization of conductivity in the quantum Hall effect. In this work, we report a similar quantization effect investigating how a perpendicular magnetic field impacts the Sublattice Charge Order (SCO) in two-dimensional Dirac insulators at charge neutrality, specifically when the Fermi energy is located strictly within the bulk energy gap. In analogy to the Ising antiferromagnetic order, where spins alternate on adjacent sites, the SCO in a bipartite lattice corresponds to an imbalance in electronic occupation between the two sublattices. Unlike a true charge density wave (CDW), which increases the unit cell, this order respects the original lattice periodicity. 

The SCO is a result of the breaking of the sublattice (inversion) symmetry leading to a gap opening in the band structure. Gapped Dirac materials with broken sublattice symmetry are of experimental and theoretical importance, especially because of their applicability in the semiconductor industry. For example, a tight-binding model of graphene with nearest-neighbor (NN) hopping does not exhibit a band gap at half-filling \cite{wallace, semenov}, but upon breaking the sublattice symmetry, a gap opens up. 
Such opening of a band gap due to the graphene-substrate interaction has been predicted theoretically and confirmed experimentally \cite{substrateinducedgraphene, grapheneonhbn, GrapheneonSic,grapheneonsicexp,stmgraphene}. Similar effects have been observed with fluorinated graphene films grown on copper foils \cite{flourine}, reversible hydrogenation \cite{hydrogenation, hydrogenationth}, and inducing sublattice potential via strain engineering \cite{Guinea2010, uniaxialDFT}.\par
However, the opening of the gap needs to be accompanied by broken sublattice symmetry to observe the SCO. The opening of a gap due to broken sublattice symmetry provides the prerequisite for a variety of interesting physical phenomena (for example, valley-contrasting physics and spin-orbit coupling \cite{review, Sun2019,Mak2018,Ma2019,Manchon2015}), and it has been well studied and observed. For example, the application of the strain induced sublattice potential to modulate the band gap in Dirac materials is a well-known method \cite{Guinea2010}. A wide range of other techniques have also demonstrated ways to manipulate inversion symmetry breaking in 2D Dirac materials \cite{Wu2013,Zhang2009,Finney2019,Sui2015}.\par

A staggered sublattice potential applies distinct onsite energies to the A and B sublattices of graphene, thereby breaking sublattice equivalence and inversion symmetry. In monolayer graphene on hexagonal boron nitride (h-BN), the inequivalence of boron and nitrogen atoms precisely induces such a potential \cite{Woods2014,Yankowitz2019,hunt}.
In addition to sublattice symmetry breaking, SCO can also result from valley symmetry breaking. The latter gives rise to important physical phenomena, such as the anomalous quantum Hall effect (QHE), first predicted in \cite{haldane}. Valley symmetry breaking also creates a tunable energy gap, another tool for band-gap engineering \cite{Hill_2011,valleyoptoelectronics,moleculardope,castro2007}.

This paper predicts and studies the magnetic field-induced quantization and quantum oscillations of SCO at charge neutrality. To this end, we consider graphene-based Dirac insulators, where the experimentally induced SCO resulting from distinct onsite potentials for the two sublattices opens a band gap. Under perpendicular magnetic fields, the massive Dirac spectrum splits into discrete Landau levels (LLs), significantly modifying the electronic density of states and consequently influencing SCO. The main results of the present paper (listed according to the strength of the external magnetic field from weak to strong) can be summarized as follows.

\begin{figure}[H]
    \centering
    \includegraphics[height=5 cm, width=7 cm]{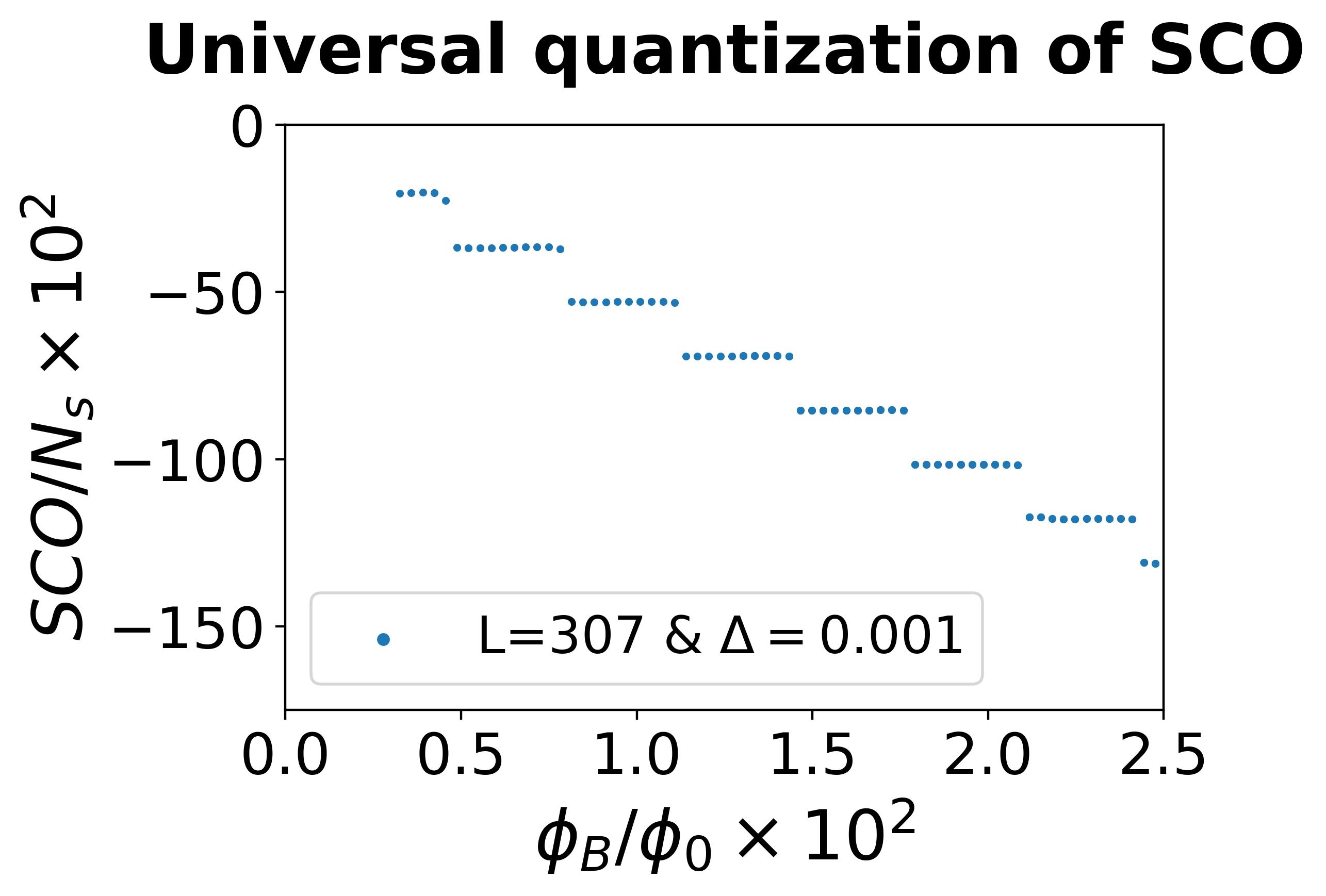}
    \caption{ Universal quantization of SCO at $\omega_c\lesssim\Delta$. Here $N_s$ is the number of sites in the finite-size system, $L$ is the system size, $\phi_B$ is the magnetic flux threading the unit cell of the honeycomb lattice, and $\phi_0$ is the flux quantum.}
    \label{fintro}
\end{figure}

\begin{enumerate}

\item At vanishingly small non-quantizing magnetic fields such that the net flux threading the system is smaller than a flux quantum, and small sublattice symmetry breaking potentials, SCO exhibits a singular magnetic field dependence stemming from next-nearest-neighbor hopping processes. 

 \item 
 In the intermediate field regime, when the cyclotron gap between the zeroth Landau level and the first Landau level is smaller than the sublattice potential $\omega_c\lesssim\Delta$, SCO exhibits universal quantized plateaus (see Fig.~\ref{fintro}) due to the discrete Landau level degeneracy. Every time the flux increases by one quantum, one electron per spin is moved from the sublattice at a higher chemical potential to the one at the lower chemical potential. This single‐electron transfer originates from the sublattice polarization of the zeroth Landau level in gapped Dirac materials, thereby realizing a topological Thouless pump \cite{CitroAidelsburger2023Thouless}.

\item 
In the strong field limit, $\omega_c\gtrsim\Delta$,  corresponding to the integer quantum Hall regime, SCO exhibits quantum magneto‐oscillations with sharp singularities at integer Landau‐level fillings, in agreement with the theory of anomalous de Haas–van Alphen oscillations in insulators without a Fermi surface when $\omega_c$ exceeds the gap.

\end{enumerate}

The reported physics of points 2 and 3 above is similar in spirit, but corresponds to different measurable characteristics and leads to different physical effects compared with the unconventional form of quantization of conductivity in graphene \cite{qhegraphene} and strong magneto-oscillations at magnetic fields, $\omega_c\gg\Delta$, in \cite{Qos, qossharapov,Panda2022Quantum}.

In subsequent sections, we systematically explore these phenomena through rigorous analytical and numerical methods, identifying explicit parameter regimes where different behaviors manifest. We propose scanning tunneling microscopy (STM) experiments \cite{E.Andrei2009,yazdani1, yazdani2, yazdani3, yazdani5} to observe predicted quantization and magnetooscillations of SCO in Dirac insulators. STM is sensitive to the local electronic density of states, and can be used to probe SCO in e.g. graphene by visualizing the local electronic structure and identifying the presence of charge density variations on the A and B sublattices. In essence, STM allows for real-space imaging of the distribution of charge carriers and the resulting ordering of electrons on the distinct carbon atoms of the graphene lattice. We show that this serves as a reliable probe of the energy gap in Dirac materials, regardless of the gap size, without performing spectroscopy.

The remainder of the paper is organized as follows. In Section \ref{hamiltonian}, we consider the Hamiltonian of the graphene monolayer with a staggered sublattice potential. We define $SCO$ and show that it vanishes in the absence of sublattice potential both in presence and absence of a magnetic field, while in the presence of sublattice potential, the $SCO$ is finite. In section \ref{Sect3}, we analytically compute $SCO$ in the presence of a magnetic field when the sublattice symmetry is broken, and derive the three regimes of weak, intermediate, and strong magnetic fields described above. In Section \ref{numerics}, we verify the analytical results predicted for weak sublattice potentials for the entire range of magnetic fields, using high precision lattice calculations via Hofstadter butterfly. 

\section{The Model and definitions}\label{hamiltonian}

Graphene at half filling is an example of a Dirac quantum material that has no Fermi surface. Its electronic structure is effectively described by a tight-binding model on a honeycomb lattice. This lattice comprises two inequivalent sublattices labeled A and B. The resulting energy dispersion hosts Dirac cones at the corners (K and K' points) of the Brillouin zone. Near these points, electrons behave as massless Dirac fermions. Figure \ref{model} provides a visual representation of the model studied in this paper. \par
The Bloch Hamiltonian of graphene with nearest-neighbor(NN) and next-nearest-neighbor (NNN) hopping in the presence of onsite sublattice staggered potential, $\Delta$, is given by equation \ref{1}:

\begin{figure}[H]
    \centering
    \includegraphics[height=4 cm, width=7 cm]{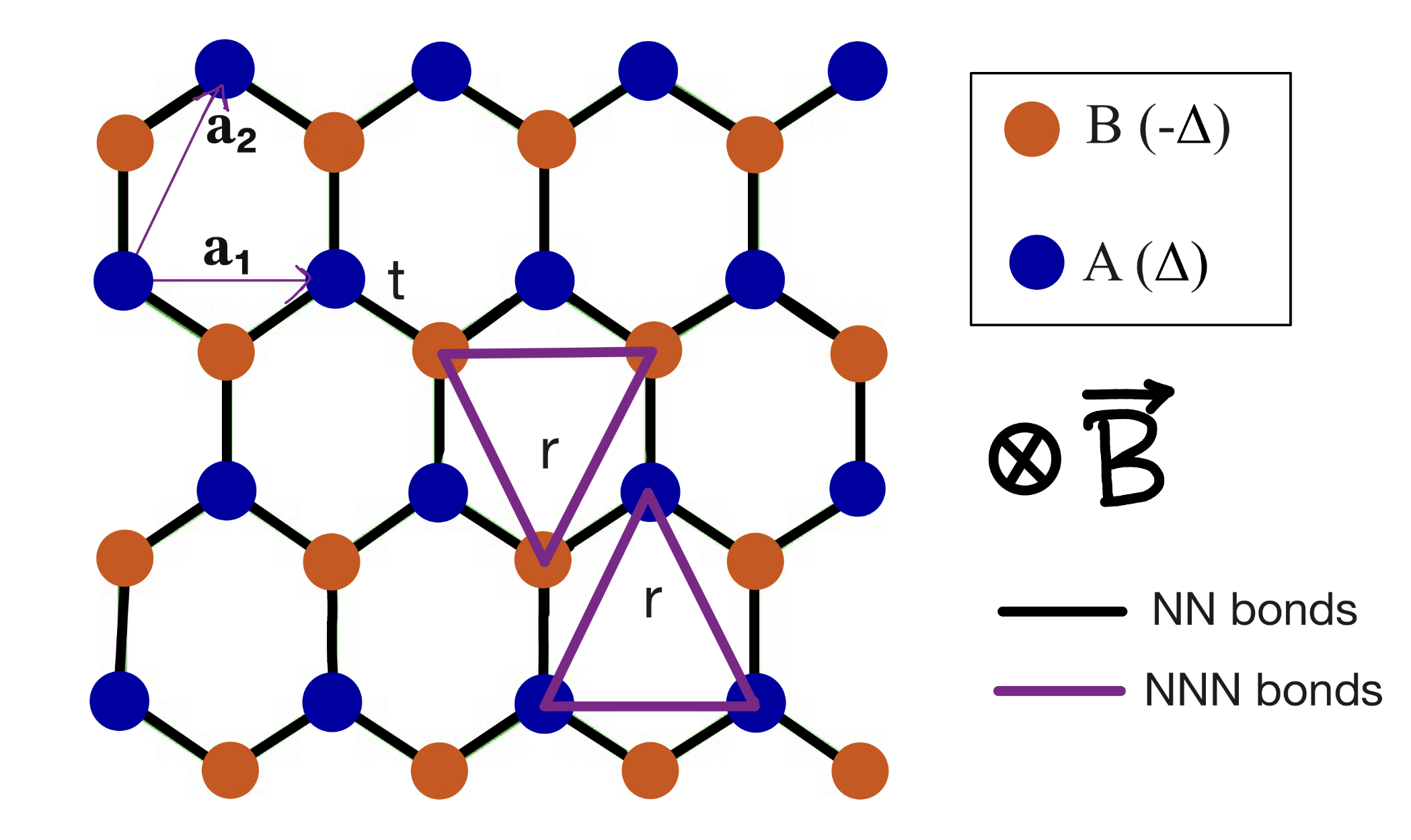}
    \caption{Tight-binding model of graphene with nearest- and next-nearest-neighbor hopping in the presence of a staggered sublattice potential, $\pm\Delta$, and an external magnetic field, ${\bf B}$.}
    \label{model}
\end{figure}

\begin{equation}\label{1}
    H (\mathbf p) = -t \begin{pmatrix}
       r T_\mathbf p^A+\Delta & T_\mathbf p\\ T_\mathbf p^* & r T_\mathbf p^B-\Delta 
    \end{pmatrix},
\end{equation}
where $t$ and $rt$ denote the NN and NNN hopping amplitudes, respectively, and 
\begin{align}
    T_\mathbf p&=e^{\ii \mathbf p\cdot\mathbf e_1}+e^{\ii \mathbf p\cdot\mathbf e_2}+e^{\ii \mathbf p\cdot\mathbf e_3}\\
    T_\mathbf p^A&=T_\mathbf p^B=e^{\ii \mathbf p\cdot \mathbf a_1}+e^{\ii \mathbf p\cdot \mathbf a_2}+e^{\ii \mathbf p\cdot \mathbf a_3}+\text{c.c.}
\end{align}
where, $\mathbf e_1=\frac{a}{\sqrt 3}(\frac{\sqrt 3}{2},\frac{-1}{2})$, $\mathbf e_2=\frac{a}{\sqrt 3}(0,1)$, $\mathbf e_3=-\frac{a}{\sqrt 3}(\frac{\sqrt 3}{2},\frac{1}{2})$ are the basis vectors connecting nearest neighbor sites of the honeycomb lattice of graphene one residing on sublattice $A$ another on sublattice $B$, and $\mathbf a_1=a(1,0)$, $\mathbf a_2=a(\frac{1}{2},\frac{\sqrt 3}{2})$, and $\mathbf a_3=a(\frac{1}{2},-\frac{\sqrt 3}{2})$ are the lattice translation vectors with $a$ representing the lattice constant. First, we shall set $\Delta=0$ and show that the system will not support $SCO$ with or without magnetic fields.  In order to study the influence of the magnetic field, we expand about the high symmetry points $K$ and $K^\prime$, in the vicinity of which the dispersion is Dirac-like. The Hamiltonian about the $K$ and $K^\prime$ is given by:
\begin{align} \label{4,5}
    \mathcal H_{K} &= \mathcal T \mathcal H(K^\prime) \mathcal T= \mathcal P \mathcal H(K^\prime) \mathcal P,\\
    \mathcal H_{K} &=-t \begin{pmatrix}
        r \bar k^2 & -\bar k_x+\ii \bar k_y \\
        -\bar k_x-\ii \bar k_y & r\bar k^2
    \end{pmatrix},
\end{align}
where, $\mathcal T$ is the time-reversal operator, $\mathcal P$ is the parity operator, $\bar p_i\equiv \sqrt 3 a p_i/2$, $\mathbf p=\mathbf k_0+\mathbf k$ and $ \mathbf k_0=K,K^\prime$ is the high symmetry point. Both Hamiltonians have the same spectrum but different wavefunctions. Diagonalizing the Hamiltonian yields two bands given by:
\begin{equation}
    \frac{E_{\bar {\mathbf k}}^\pm}{-t}=r \bar{\mathbf k}^2\pm\sqrt{{\bar{\mathbf k}^2}},
\end{equation}
where, $\mathbf k=(k_x,k_y)$ is the eigenvalue corresponding to the momentum operator.

The corresponding normalized eigenstates, namely the real-space Bloch wavefunctions for a single particle on a honeycomb lattice can be solved from the eigenvalue equation:

\begin{equation}\label{7}
    \mathcal H_K(\mathbf k)\Psi_K^\mathbf k=E_\mathbf k \Psi_K^\mathbf k,\qquad\Psi_K=\mathcal P \Psi_{K^\prime}
\end{equation}

The energy is independent of the valley index because the valley degeneracy isn't lifted (and we won't lift it throughout the paper). The wavefunction is of the form (for the exact expression see appendix \ref{A}):
\begin{equation}
    \Psi_K^\mathbf k(\mathbf x) =\begin{pmatrix}
        \psi_{A,K}^\mathbf k ({\bf x}) \\ \psi_{B,K}^\mathbf k({\bf x)}
    \end{pmatrix}
\end{equation}
 reflecting the two-sublattice nature of the honeycomb lattice.
 
To this end, we define sub-lattice charge order, $\Lambda$, the order parameter to measure the charge imbalance between two sublattices, as follows:
\begin{align}\label{91011}
    \Lambda_\eta^\mathbf k(\mathbf x)&=(\Psi_\eta^\mathbf k)^\dagger\sigma_z\Psi_\eta^\mathbf k=|\psi_{A,\eta}^\mathbf k({\bf x})|^2-|\psi_{B,\eta}^\mathbf k({\bf x})|^2\\
    \Lambda_\eta^\mathbf k &=\int d^2 \mathbf x\, \Lambda_\eta^\mathbf k(\mathbf x)\\ \Lambda&=\sum_{\eta}\sum_{\mathbf k} \Lambda_\eta^\mathbf k
\end{align}
where, we will be referring to $\Lambda_\eta^\mathbf k(\mathbf x)$ as the SCO density and $\eta=\pm1$ corresponds to $K$ and $K^\prime$ valley respectively. In the absence of a magnetic field, the SCO vanishes identically at both valleys, $\Lambda_K^{\mathbf k}=\Lambda_{K^\prime}^{\mathbf k}=0$ (see Appendix \ref{A} for details). This result follows directly from sublattice symmetry: there is no energetic preference for occupying sublattice A versus B, and thus no imbalance can develop. By contrast, when a magnetic field is applied, the SCO at each individual valley becomes finite. Nevertheless, the underlying sublattice symmetry relates the Hamiltonians at $K$ and $K^\prime$, forcing their contributions to cancel so that the net SCO remains zero:

\begin{equation} \label{12}
    \mathcal H(K^\prime)=\sigma_x \mathcal H(K) \sigma_x.
\end{equation}

The relation $\Lambda^{n,k}_K+\Lambda_{K^\prime}^{n,k}=0$ 
remains valid even in the presence of a magnetic field. Here, $\Lambda_\eta^{n,k}$ represents the sublattice charge order (SCO) associated with a Landau level (LL) characterized by the integer index $n$ and momentum $k$ (see Appendix \ref{D} for the derivation of LLs). This result is expected because the magnetic field does not change the fact that there is no additional energy cost for occupying either sublattice when $\Delta = 0$ (see Appendices \ref{C} and \ref{B}).
Moreover, the result is independent of the next-nearest-neighbor (NNN) hopping amplitude $r$. Although the Hamiltonian formally includes a finite $r$, the conclusion holds for any $r$, meaning both with and without NNN hopping. This robustness follows from the fact that NNN hopping preserves sublattice symmetry and therefore cannot induce SCO, regardless of the presence of a magnetic field.

\section{sublattice charge order}
\label{Sect3}

In the following subsections, we systematically examine the SCO in Dirac insulators, using graphene with a staggered chemical potential introduced in the previous section as an example. We start with the case of broken sublattice symmetry without a magnetic field. Next, we explore the effect of an external magnetic field, illustrating the quantization of SCO at intermediate field strengths and the magneto-oscillations of SCO in strong fields. We also consider the effects of NNN hopping and, finally, the influence of electron doping, which moves the system away from half-filling.

\subsection{SCO in Graphene with broken sublattice symmetry in absence of magnetic field}\label{scowihtoutB}

We have established that SCO vanishes as long as sublattice symmetry is preserved. In a zero magnetic field, the SCO in each valley is zero, while in a finite field, the contributions from the two valleys cancel (see Eq.~\ref{12}). If, however, the Hamiltonian contains an interaction term $\hat V(\mathbf k)$ that does not commute with $\sigma_x$,  $[V(\mathbf k),\sigma_x]\neq 0$, then sublattice symmetry is broken and Eq.~\ref{12} no longer holds. This leads to a finite SCO with a nontrivial magnetic-field dependence. The simplest realization is Eq.~\ref{13}, a graphene Hamiltonian with a staggered onsite potential $\Delta$. Physically, such a potential favors occupation of one sublattice over the other, naturally producing a nonzero SCO.

The staggered sublattice potential breaks sublattice and inversion symmetry, yielding a nonzero SCO both with and without a magnetic field. This potential is represented as $-\Delta t \sigma_z$, so the valley Hamiltonian is modified accordingly:
\begin{equation}\label{13}
    \mathcal H_\eta=-t \begin{pmatrix}
        r \bar k^2+\Delta & -\eta\bar k_x+\ii \bar k_y \\
        -\eta\bar k_x-\ii \bar k_y & r\bar k^2-\Delta
    \end{pmatrix},
\end{equation}
where, $\eta=\pm 1$ corresponds to the $K$ and $K^\prime$ valleys. The energies are modified due to the presence of onsite potential, however, the staggered potential does not lift the valley degeneracy \cite{haldane,semenov}. At zero temperature and half-filling (we only consider this scenario in the present paper) the SCO can be computed exactly using the wavefunctions corresponding to the negative energy band (see appendix \ref{A} for details) and it is given by:
\begin{equation}
    \sum_\eta\Lambda_\eta^\mathbf k=-\frac{2\Delta}{\sqrt{\bar{k}^2+\Delta^2}},
\end{equation}
where we used $\bar k\equiv |\bar{\mathbf k}|$ throughout the paper. One can now compute the total SCO across the graphene sample at half-filling, where all the states of  the Dirac sea are filled:
\begin{equation}
    \Lambda=\sum_\mathbf k\sum_\eta \Lambda_\eta^\mathbf k= \int \frac{-2\Delta}{\sqrt{\bar{\mathbf k}^2+\Delta^2}}\frac{A}{2\pi}k\, dk,
\end{equation}
where $A$ is the area of the sample. We have to recall that all the calculations and results that we have so far are about small fluctuations around the high symmetry point. If one departs from the high symmetry points, then the spectrum is no longer described by the linear dispersion at large momenta $\bar{k}\gg \Delta$: $\frac{E_k^\pm}{-t}=\pm\sqrt{\bar{k}^2+\Delta^2}\rightarrow \pm \bar{k}$ or, in other words, one cannot ignore the curvature in the dispersion relation for a large momentum. So, one has to introduce a cutoff energy $-\alpha t$ up to which the linear approximation for the dispersion works, and hence one will have an ultraviolet cutoff momentum $k_c$ defined by:
\begin{equation}
    \alpha=\sqrt{\bar k_c^2+\Delta^2}.
\end{equation}
This gives the SCO contribution at half filling originating from the Dirac spectrum:
\begin{equation}
    \Lambda=-\frac{4\Delta}{3\pi}\frac{A}{a^2}\alpha.
\end{equation}

The above result is consistent if $\Delta\ll\alpha$, and we will show that this indeed holds in what we will be referring to as the weak-sublattice potential regime. Within this assumption,  the expression for SCO per site at half-filling can be cast in the form:
\begin{equation} \label{sco0}
    \frac{\Lambda}{N_s}=-\frac{\Delta}{\sqrt 3 \pi}\alpha,
\end{equation}
where $N_s$ denotes the total number of sites. When there is no magnetic field, the SCO does not depend on the NNN hopping parameter $r$. We will later demonstrate that finite $r$ introduces only subleading field-dependent corrections to the SCO.

For energies up to the ultraviolet cutoff $\alpha = 0.46$, the exact lattice and Dirac spectra agree within 7\% accuracy. This cutoff value is independently confirmed through lattice calculations in a magnetic field (see Fig.~\ref{f2}). An important question is what fraction of particles occupy the Dirac portion of the spectrum? We first evaluate this without next-nearest-neighbor (NNN) hopping and then show that NNN corrections are subleading. To this end, we denote by $N_{DS}$ the number of particles filling the Dirac sea, which can be estimated from the following relation: $N_{DS}=2\sum_{\varepsilon_\mathbf k<\alpha}1$. The extra factor of 2 comes from the two valleys that contribute to the SCO. This gives us the relation between the number of sites $N_s$ and $N_{DS}$: $N_{DS}=\frac{\alpha^2}{2\sqrt 3\pi}N_s$ (for $r=0$). At half-filling, the total number of particles $N$ is equal to $N_s/2$, and thus
\begin{equation}
    \frac{N_{DS}}{N}=\frac{\alpha^2}{\sqrt3 \pi}\simeq0.04.
\end{equation}
This shows that only a small number of particles occupy the Dirac sea. However, we show that the Dirac sea completely explains the SCO at intermediate magnetic fields corresponding to $\omega_c\lesssim \Delta$, where $\omega_c$ is the cyclotron gap between the zeroth Landau level and the first Landau level. This implies that at these intermediate magnetic fields, the contribution to the SCO coming from particles outside the Dirac sea is negligibly small, and hence our analytical analysis of the SCO becomes essentially exact in this regime. 

It is interesting to consider the effects of NNN hopping on the number of particles occupying the Dirac sea. In the limit where the Dirac spectrum is more dominant than the NNN interactions, i.e. $r\alpha\ll1$, the effect is subleading and is given by:
\begin{equation} \label{nds}
    N_{DS}=N_s\frac{\alpha^2}{2\sqrt 3\pi}(1-2r\alpha).
\end{equation}
It is important to point out that the typical values of the NNN hopping parameter $r$ in units of the NN hoping parameter $t$ is $r\sim 0.1$ \cite{reichnnn,electronicpropofgraphene, Rozhkov2016, nnnexp}. This is well within our assumption $r\alpha\ll1$, implying that the Dirac sea has the dominant contribution to the SCO even in the presence of NNN hopping.

Away from charge neutrality, the result is modified and becomes dependent on the Fermi-energy, $\epsilon_F=-t\,\varepsilon_F$, as
\begin{equation}
    \frac{\Lambda_{\nu>\frac{1}{2}}}{N_s}=-\frac{\Delta}{\sqrt 3 \pi}(\alpha-\varepsilon_F).
\end{equation}
Here, we have assumed $\varepsilon_F\lesssim\alpha$, that is, the ultraviolet cutoff is the largest energy scale in the low-energy description of the model (which is still much smaller than the bandwidth defined in terms of the NN hopping parameter, $t$). Moreover, one can find the particle density behavior of the SCO. The asymptotic behavior of SCO at low and high densities is presented in appendix \ref{C}.

\subsection{Quantization of SCO}\label{2a}

We have established that sublattice symmetry breaking is essential for realizing SCO. In earlier sections, we derived its behavior at zero magnetic field and showed that it is independent of the NNN hopping term $r$. Here, we examine SCO in the presence of a magnetic field under a weak sublattice potential, $\Delta \ll \alpha$, in the regime 
$\omega_c \lesssim \Delta$. In this case, SCO takes the form

\begin{equation}\label{222}
    \frac{\Lambda}{N_s}=\frac{-\Delta\alpha}{\sqrt 3\pi}-\frac{d}{N_s}+\frac{1}{\sqrt 3}r\phi_b \alpha^2/\Delta,
\end{equation}
where $d=\phi_{\text{tot}}/\phi_0$ represents the Landau-level degeneracy, with $\phi_{\text{tot}}$ being the total flux and $\phi_0$ the flux quantum, while $\phi_b = \sqrt{3}e\phi_B/2$ and $\phi_B$ is the flux through a unit cell.

The first term, surviving in the $B=0$ limit, reproduces the zero-field result. The second term represents SCO quantization in this regime, originating from the zeroth Landau level and being universal across Dirac materials. The third term accounts for NNN hopping corrections, which are subleading compared to the first two. Both the first and third terms depend on the UV cutoff $\alpha$ of the Dirac spectrum and are therefore non-universal, varying with material specifics.

We now derive Eq. \ref{222}, first excluding NNN hopping, and then incorporate its corrections in the presence of a staggered sublattice potential, where energetic preference favors occupation of one sublattice over the other. The Hamiltonian of interest is Eq.~\ref{1} with $r=0$. The staggered sublattice potential, $\Delta$, breaks the inversion symmetry and induces the sublattice charge order. In a magnetic field, the eigenstates are Landau wavefunctions $\Phi_{n,k}$ (see Appendix \ref{E}), where $n$ is the Landau-level index and $k$ labels the degenerate states within each level (translation quantum number along $x$). In the Landau gauge, eigenstates are specified by $(n,k)$ rather than $\mathbf k$, so the definition of SCO in Eq.~\ref{91011} is modified accordingly:

\begin{align}
    \Lambda_\eta^{(n,k)}(\mathbf x)&=(\Psi_\eta^{(n,k)})^\dagger\sigma_z\Psi_\eta^{(n,k)}\nonumber\\
    &=|\psi_{A,\eta}^{(n,k)}({\bf x})|^2-|\psi_{B,\eta}^{(n,k)}({\bf x})|^2\\
    \Lambda_\eta^{n,k} &=\int d^2 \mathbf x\, \Lambda_\eta^{n,k}(\mathbf x)\\ \Lambda&=\sum_{\eta}\sum_{{n,k}} \Lambda_\eta^{n,k}
\end{align}

After linearizing the spectrum around the high symmetry points, the magnetic field-dependent contribution to SCO is given by (see Appendix \ref{E} for details):
\begin{equation}
    \Lambda_{n,k}= \frac{1-c_n^2}{1+c_n^2}+\frac{c_n'^2-1}{1+c_n'^2}
\end{equation}
where, $c_n=-\sqrt{1+\Delta_n^2}-\Delta_n$, $c_n'=-\sqrt{1+\Delta_n^2}+\Delta_n$, 
$\Delta_n^2=\frac{\Delta^2}{2\phi_bn}$. We have already added the contribution from both valleys in the above equation and integrated over the sample (since the Landau wavefunctions are normalized, the integrals are just equal to $1$). The SCO density for each Landau level is localized near $y_0=k l_b^2$ and is uniform along the $x$ direction due to our choice of gauge. If we switch to a different gauge, the SCO density changes (for example, in the symmetric gauge, it will be localized within a radius dictated by the angular momentum), however, the total SCO across the sample is a gauge invariant quantity.
In order to compute the SCO contribution at half-filling, we need to sum over the Landau levels and momenta. The sum over $k$ will give the integer degeneracy $d=[\phi_{tot}/\phi_0+0.5]$, where, $[\ldots]$ represents smallest integer function and $\phi_{tot}=B A$ is the flux through the sample ($\sum_k\xrightarrow{}d$):
\begin{equation}\label{24}
    \Lambda=d\sum^{N_{LL}}_{n=1}\left( \frac{1-c_n^2}{1+c_n^2}+\frac{c_n'^2-1}{1+c_n'^2}\right) + \Lambda_0,
\end{equation}
where $\Lambda_0$ is the contribution from the zeroth LL and, $\phi_b=\sqrt 3 e \phi_{B}/2$ with $\phi_{B}$ being the magnetic flux through the unit cell, $N_{LL}$ is the number of filled Landau levels per valley which is estimated from
\begin{equation}
    \sqrt{2 N_{LL} \phi_b+\Delta^2}\sim\alpha
\end{equation} 
giving $N_{LL}=\alpha^2/2\phi_b$ when $\Delta<\alpha$. Here, we introduced a new notation $\phi_b$ for convenience as it is related to the cyclotron gap, $\omega_c$, between the zeroth and first LL by $\omega_c=\sqrt{2\phi_b+\Delta^2}-\Delta\simeq\phi_b/\Delta$.

We shall now calculate the contribution from the zeroth LL. The wavefunction for zeroth LL at $K$ valley is given by $\Psi_K^{0,k}=(
    0,\,\,\,\,\, \Phi_{0,k}
)^t$ and has energy $-\Delta$. The zeroth LL at the $K^\prime$ valley has energy $+\Delta$ and hence will not contribute at half-filling. Therefore, the contribution to $\Lambda$ due to the zeroth LL is $\Lambda_0=-d$. In the absence of the onsite potential, the zeroth LLs of the two valleys have zero energy, and their SCOs cancel each other out, just like the other LLs. Combining the contribution from zeroth LL and simplifying the terms in equation \ref{24}, the sum reduces to:

\begin{equation}
    \Lambda=-2\,d\sum_{n=1}^{N_{LL}}\sqrt{\frac{x}{2n+x}}+\Lambda_0,
\end{equation}
where $x=\Delta^2/\phi_b$ which is the ratio of sublattice potential and cyclotron frequency, $\omega_c$, i.e. $x=\Delta/\omega_c$. We shall convert the sum into an integral using the integral representation of Gamma function as follows:
\begin{equation}
    \Lambda-\Lambda_0=\frac{-2d}{\Gamma(1/2)}\sqrt{\frac{x}{2}}\sum_{n=1}^{N_{LL}}\int^\infty_0 dt\,t^{-1/2}e^{-t}\frac{1}{\sqrt{n+x/2}},
\end{equation}
representing the contribution to SCO from all LLs except the $n=0$. Here, the $n$ dependence can be absorbed into the exponential by the change of variable: $z=N_{LL}t/(2n+x)$. This converts the sum in $n$ into a geometric series which can be evaluated exactly, giving the following integral representation for SCO:
\begin{align}
    \Lambda-\Lambda_0&=-\frac{2d}{\Gamma(\frac{1}{2})}\sqrt{\frac{x}{N_{LL}}}\nonumber\\&\times\int^\infty_0 \frac{dz}{\sqrt z}\,e^{-z(1/N_{LL}+x/N_{LL}+1)} \frac{\sinh{z}}{\sinh{(z/N_{LL})}}
\end{align}

We shall analyze this integral at $x\gtrsim 1$, i.e. at weak and intermediate magnetic field strengths, when the cyclotron frequency is smaller than the sublattice potential. At these magnetic fields, $N_{LL}$ is a large number, so $1/N_{LL}\ll 1$. Furthermore, using the definitions of $x$ and $N_{LL}$, we have $x/N_{LL}=2\Delta^2/\alpha^2\ll1$ when $\Delta\ll\alpha$. Using these properties, we can now estimate the contributions to the integral coming from two integration regions $z\ll N_{LL}$ and $z\gg N_{LL}$ and confirm the estimates numerically.

When $z\ll N_{LL}$, one can approximate the exponential term by $e^{-z}$ and $\sinh{z/N_{LL}}\sim z/N_{LL}$, yielding a contribution 
\begin{equation}
    \Lambda-\Lambda_0\simeq-\frac{2 d}{\Gamma(\frac{1}{2})}\sqrt{x N_{LL}}\int^\infty_0 \frac{dz}{\sqrt z}e^{-z}\frac{\sinh z}{z}
\end{equation}
The integral is equal to $\sqrt{2\pi}$, giving $\Lambda-\Lambda_0=-2d\sqrt{2xN_{LL}}$. Substituting the expressions for $x$ and $N_{LL}$, the SCO can be cast in the following form:
\begin{equation}
    \Lambda-\Lambda_0=-
    \frac{4 \Delta \alpha}{3\pi}\frac{A}{a^2}.
\end{equation}
The latter simplifies to the answer we obtained from the zero magnetic field calculation in equation \ref{sco0}:
\begin{equation}
    \Lambda-\Lambda_0=\frac{-\Delta\alpha}{\sqrt 3\pi} N_s.
\end{equation}

The contribution of $z\gg N_{LL}$ is subleading as the integrand is exponentially suppressed by $e^{-z \,x/N_{LL}-2z/N_{LL}}$ and, hence, can be neglected.
Combining the two results we have the following result of SCO at magnetic fields corresponding to $\omega_c\lesssim \Delta$:
\begin{equation}
    \frac{\Lambda-\Lambda_0}{N_s}=\frac{-\Delta\alpha}{\sqrt 3\pi},
\end{equation}
Substituting the expression for the zeroth LL, one obtains
\begin{equation}\label{33}
    \frac{\Lambda}{N_s}=\frac{-\Delta\alpha}{\sqrt 3\pi}-\frac{d}{N_s}.
\end{equation}
Therefore, in this regime, the SCO is expected to be universally quantized, increasing in increments of $-1/N_s$ (referred to as plateaus), entirely arising from the zeroth LL when the degeneracy changes. This aspect of the result originates solely from the zeroth LL and is \textit{universal} to all Dirac materials. The outcome for weak magnetic fields (when the flux through the system is smaller than a flux quantum) represents the first plateau. This is not surprising, as there is no LL quantization in weak fields (since the gap between successive levels vanishes). Consequently, the result resembles that of a zero field. This aspect is \textit{non-universal} and depends on the lattice or model considered; the prefactors and cutoff rely on these factors. Thus, the SCO at magnetic fields, $\omega_c\lesssim \Delta$, has two components. The first term represents the SCO at zero magnetic field and is \textit{non-universal}, i.e., model-dependent. The second component leads to \textit{universal} quantization of the SCO in Dirac materials. Although the steps exist only in the presence of a sublattice potential, the gap between the plateaus is independent of that potential (gap in the spectrum), representing single electron transfer between sublattices per flux quantum, resulting from the sublattice polarization of the zeroth Landau level in gapped Dirac materials. Such a topological Thouless pump effect for SCO can detect the energy gap in Dirac materials, irrespective of the gap size. Later, these results will be verified using lattice calculations via the Hofstadter Butterfly in section \ref{numerics}. 

\subsection{Effects of NNN hopping}

As shown in Sec.~\ref{scowihtoutB}, SCO is unaffected by NNN hopping in the absence of a magnetic field. With a magnetic field, NNN hopping contributes only at vanishingly small fields, $\pi\phi_{tot}/\phi_0 \ll \Delta^2/\alpha^2$, in the presence of a weak sublattice potential. This yields a subleading, nonuniversal correction to the SCO plateau of order $\sim r\Delta |B|$, which disappears beyond this regime and does not alter the universal quantization of SCO at intermediate fields [Eq.~\ref{222}]. The $r$-dependent term in Eq.~\ref{222} is model-specific, and the derivation with relevant scales is provided in Appendix~\ref{NNN section}.

Thus, the only experimentally accessible field-dependent feature that is sensitive to the NNN hopping is the singular field dependence of the first, nonuniversal plateau of the SCO. At stronger magnetic fields, this subleading NNN-related correction disappears, leaving no measurable trace. Detecting the singular field dependence of the first nonuniversal plateau requires experiments on finite-size systems subjected to extremely weak magnetic fields. While STM is, in principle, sensitive to such small corrections, identifying the singularity demands an additional step: one must extract the derivative of the experimentally measured SCO with respect to the magnetic field and observe the discontinuity of this derivative at $|B|\rightarrow 0$. This makes the experimental verification of the effect technically demanding.

\subsection{Beyond half-filling}

Beyond half-filling at chemical potentials more than the sublattice potential, $\varepsilon_f>\Delta$, the contribution to SCO from the zeroth LL vanishes since the SCO from the zeroth LL coming from electrons in the vicinity of  $K^\prime$  cancels out the contribution from the $K$ valley. The SCO is therefore given by:
\begin{equation}
\Lambda_{\nu>1/2}=-2\,d\sum_{n=n_F}^{N_{LL}}\sqrt{\frac{x}{2n+x}},
\end{equation}
where $n_F$ is the LL corresponding to the Fermi energy i.e., $\sqrt{2n_F\phi_b+\Delta^2}=\varepsilon_F$. Then, it is convenient to write the sum as $\sum_{n=n_F}^{N_{LL}}=\sum_{n=1}^{N_{LL}}-\sum_{n=1}^{n_F}$. This allows one to apply the result of the previous section to immediately see that:
\begin{equation}\label{ff}
    \frac{\Lambda_{\nu>1/2}}{N_s}=\frac{-\Delta\left(\alpha-\varepsilon_F\right)}{\sqrt 3\pi},
\end{equation}
which is the dominant term that does not depend on the magnetic field. Eq.~(\ref{ff}) corresponds to the SCO in the range of intermediate fields to the zero magnetic field limit.

\begin{figure}[H]
    \centering
    \includegraphics[scale=0.23]{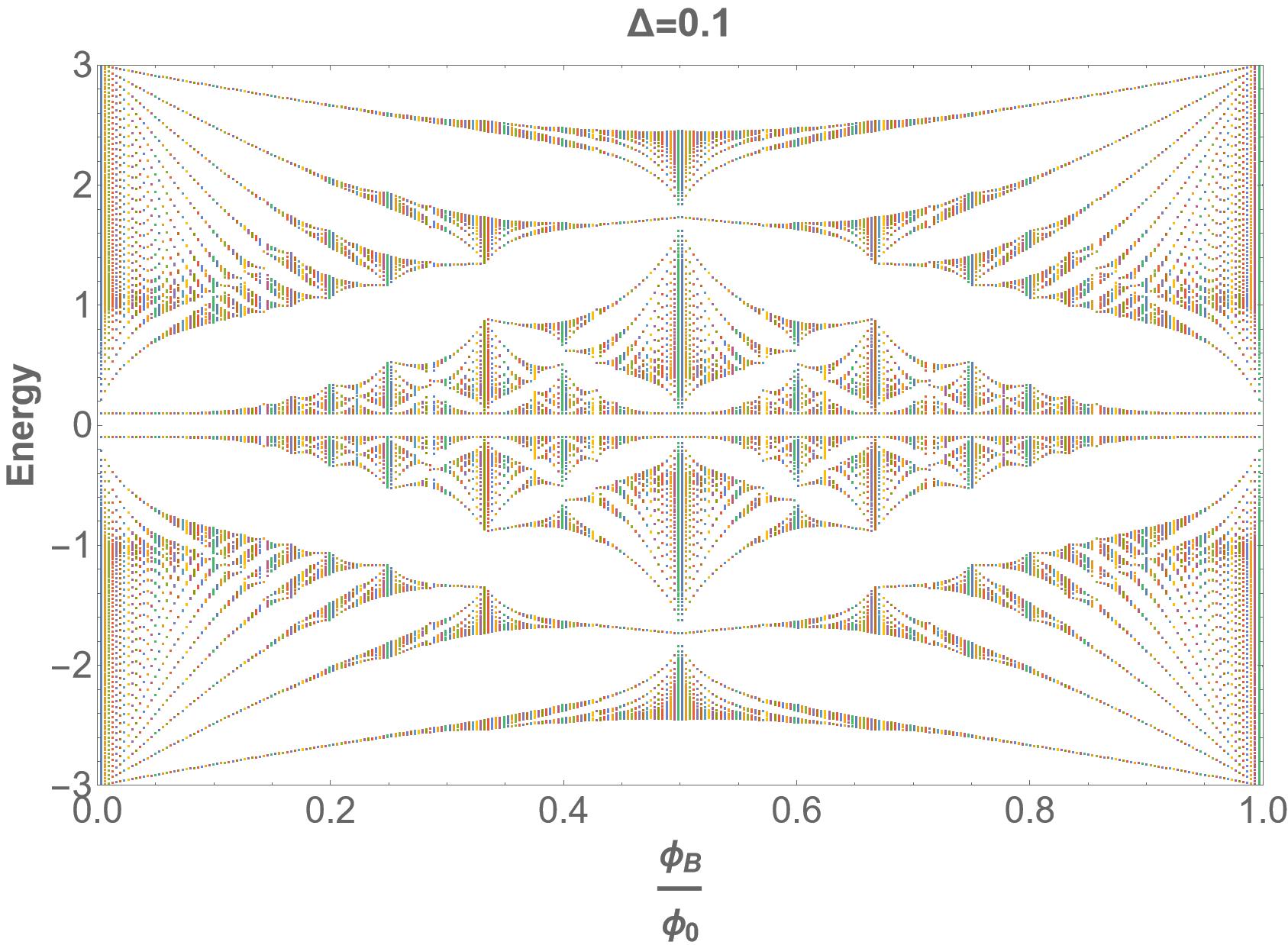}
    \caption{Energy spectrum of lattice  electrons in graphene with sublattice potential, $\Delta=0.1$ (spectral gap measured in the units of NN hopping). Here $\phi_B$ is the flux through the unit cell and $\phi_0$ is the flux quantum.}
    \label{f1}
\end{figure}

We would like to emphasize again that this result is only valid at weak and intermediate magnetic fields, $\phi_b\ll\Delta^2$, which correspond to $\omega_c\ll\Delta$, and weak sublattice potentials ($\Delta\ll \alpha$) regime. Outside of this regime, beyond the UV cut-off, the SCO contribution comes beyond the Dirac spectrum and can only be captured by lattice calculations.  Using the Hofstadter butterfly spectrum in section \ref{numerics}, we will estimate these regimes from the lattice calculations. We will resort to lattice calculations to study SCO outside this regime, particularly for observation of quantum oscillations, based on the simulation of the Hofstadter Butterfly spectrum.

In the next section, we will confirm our analytical calculations using the exact calculation of the SCO from the Hofstader butterfly spectrum. We will also investigate the strong field limit and observe quantum oscillations.

\begin{figure}[H]
    \centering
    \includegraphics[scale=0.24]{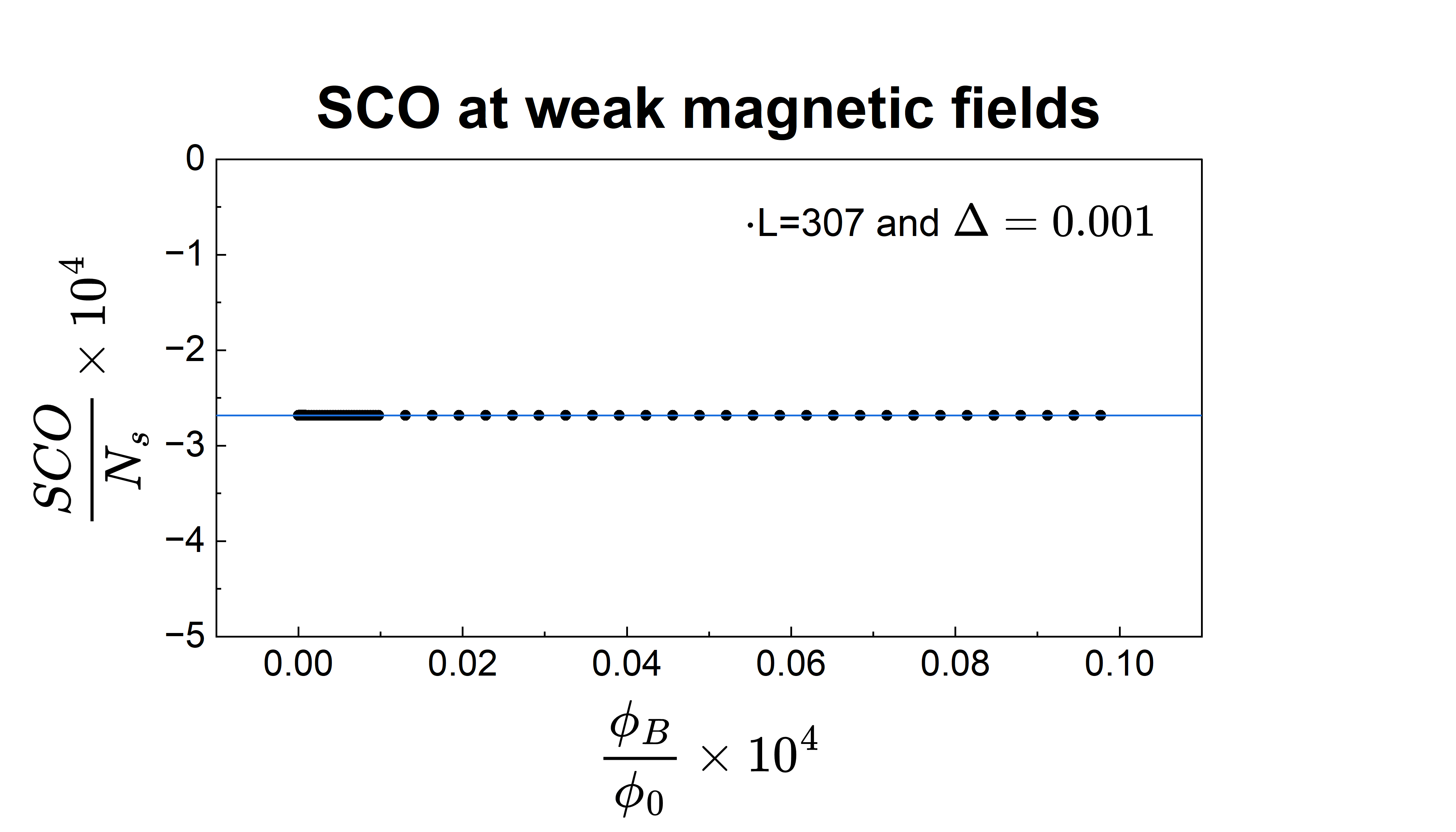}
    
    \caption{$SCO$ per site obtained from the Hofstadter butterfly showing the nonuniversal plateau at weak magnetic fields such that the net flux threading the system is smaller than a flux quantum. The plateau value is related to the cutoff $\alpha$ via Eq.~\ref{sco0}, resulting in the cutoff value in Eq.~\ref{afa}.}
    \label{f2}
\end{figure}

\section{Hofstadter Butterfly Lattice Calculations}\label{numerics}

As discussed in the previous sections, the SCO is quantized at magnetic fields corresponding to a cyclotron frequency less than the spectral gap (the same as the condition $\phi_b\lesssim \Delta^2$). We start this section by self-consistently finding the value of the cutoff $\alpha$ with high precision from the Hofstadter Butterfly. In this section, we shall only consider the spectrum of Graphene with sublattice potential at half-filling.  Furthermore, we will also study the $SCO$ per site across different magnetic field strengths. We will confirm our analytical calculations of SCO observing three distinct magnetic field regimes for which  SCO shows qualitatively different behaviors.

\begin{figure}[H]
    \centering
    \includegraphics[height=8.5cm, width=8.5cm]{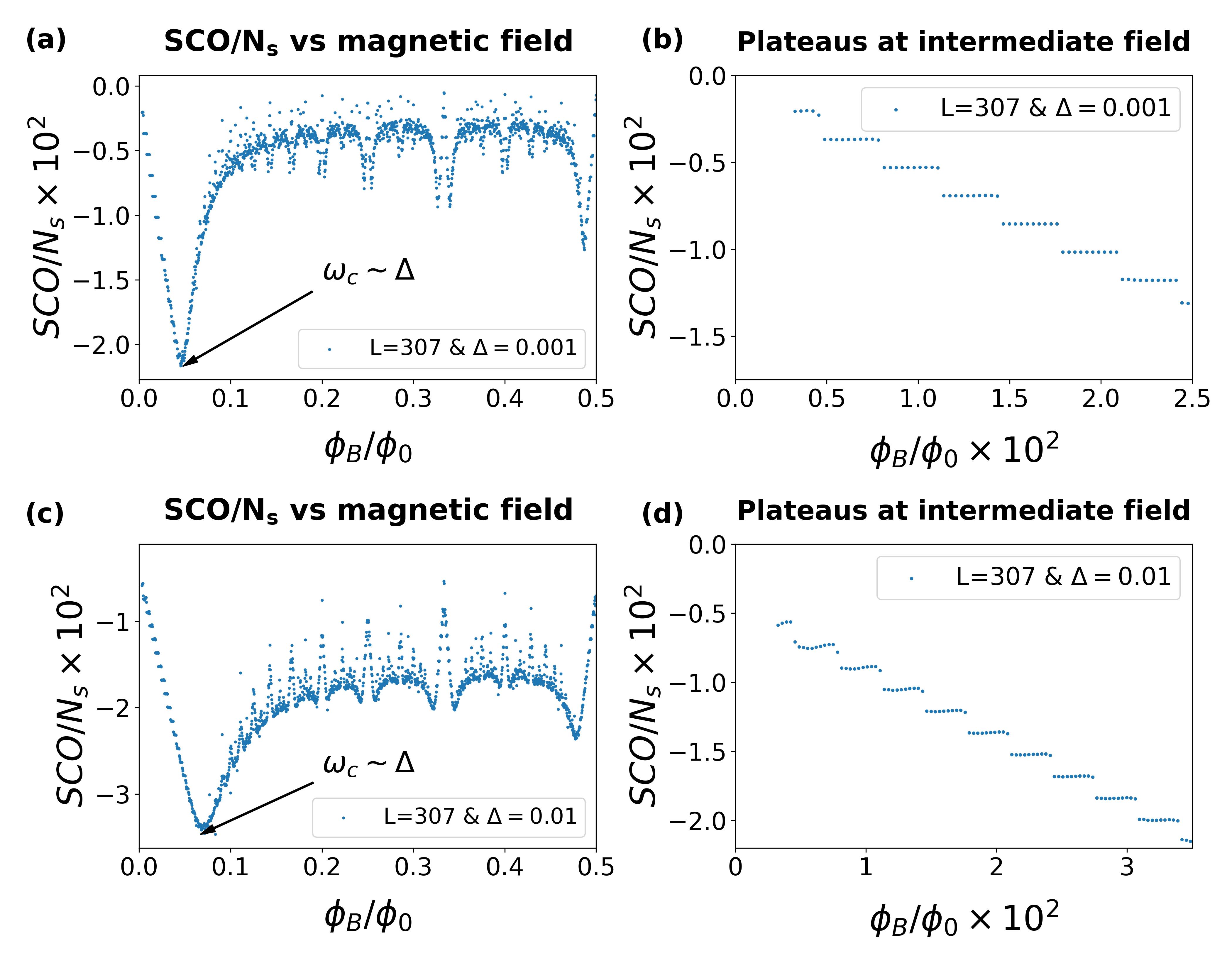}
    
    \caption{The behavior of SCO per site is shown across different regimes of the external magnetic field. Panels (a) and (c) display the SCO over the full range of the magnetic field, from weak to strong fields separated by the deep minimum corresponding to $\omega_c=\Delta$. At strong fields corresponding to $\omega_c>\Delta$, SCO exhibits magneto-oscillatory behavior. In contrast, panels (b) and (d) provide a magnified view of the intermediate-field regime, where the SCO per site exhibits universal quantization.}
    \label{3}
\end{figure}
These are the weak (total magnetic flux is smaller than $\Phi_0$), intermediate ($\omega_c\lesssim\Delta$), and strong quantizing ($\omega_c\gtrsim\Delta$ field regimes. In the weak field and weak sublattice potential regime, the exact SCO from the Hofstadter butterfly spectrum exactly matches our analytical results,  Eq.~(\ref{33}), where $SCO/N_s$ is universally quantized, forming plateaus. The spacing between the plateaus also matches our analytical results exactly, confirming one electron transfer between sublattices per flux quantum, which results from the sublattice polarization of the zeroth Landau level.
At strong fields, the $SCO$ shows strong magneto-oscillations based on anomalous De Haas-Van Alphen oscillations in insulators reported in \cite{Qos}. We would like to remind the reader that the calculations in this section are done at half-filling and zero temperature.

To study the spectrum and SCO in a magnetic field, we use a lattice Hamiltonian and exact diagonalization method. The electron spectrum in a magnetic field is given by the Hofstadter butterfly (Fig. \ref{f1}), with gaps opened by the staggered sublattice potential. The system sizes employed in exact diagonalization for the calculation of SCO are specified in the corresponding figure captions. Solving the Hofstadter problem yields both the exact spectrum and wavefunctions, from which the SCO can be computed directly. 

Below, we verify the analytical results in the weak magnetic field and weak-sublattice potential regime. At the next step, we will derive our main result, i.e., the quantization of SCO in the intermediate magnetic field regime at weak-sublattice potentials. Finally, we will discuss the results of the de Haas-van Alphen regime. The lattice calculations have been bench-marked by investigating the dependence of cut-off on system size and sublattice potential in appendix \ref{benchmark}.

\begin{figure}[H]
    \centering
    \includegraphics[height=8.5cm, width=8.5cm]{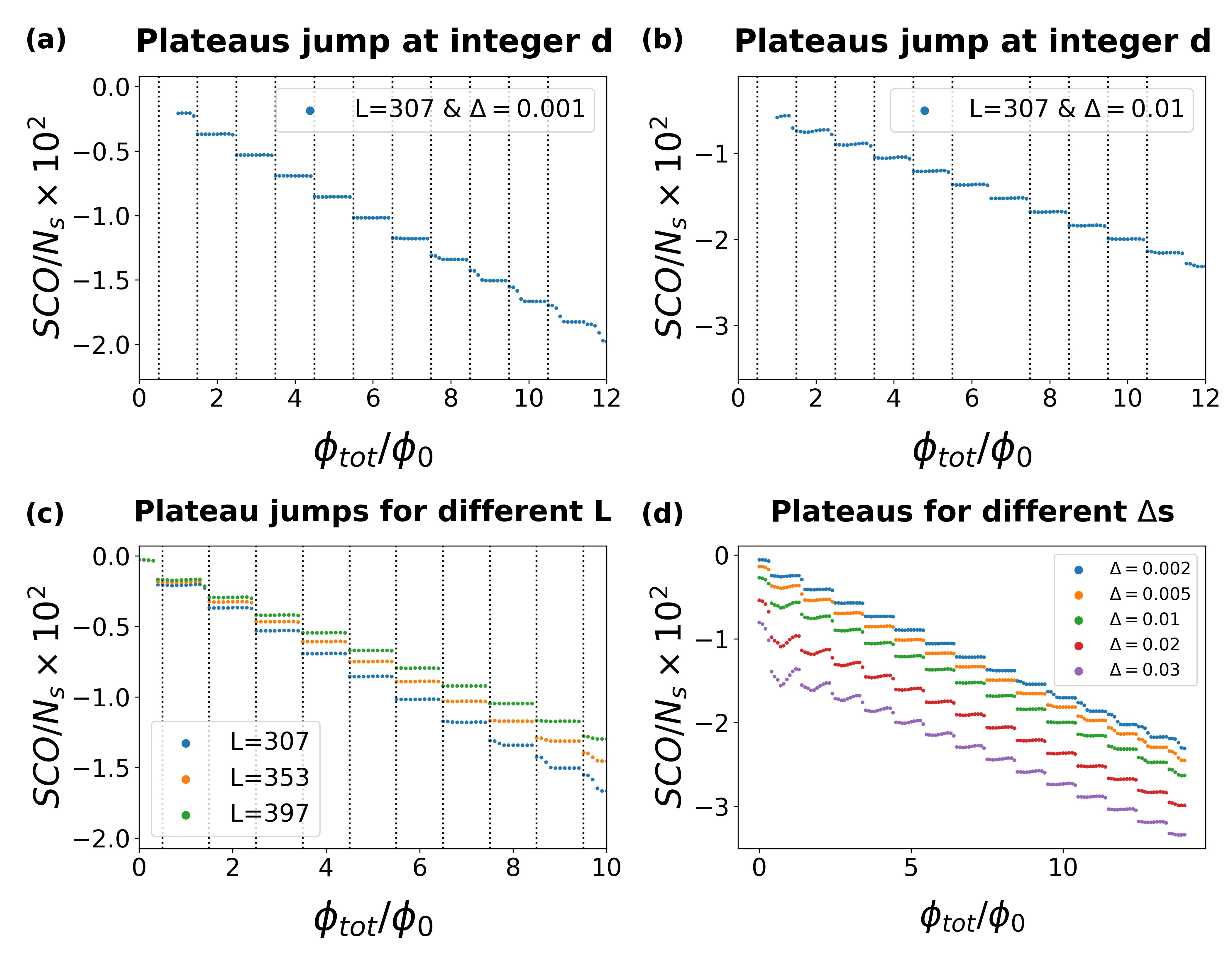}
    
    \caption{SCO per site versus the total flux through the sample, $\phi_{tot}$, in the units of the flux quantum, $\phi_0$. The plateau transitions align with the vertical lines in panels (a)–(c), which occur at $\phi_{tot}/\phi_0 = n + 0.5$ ($n = 1,2,3,\dots$), equivalently at integer values of $d \in \mathbb{Z}+$. Panel (d) shows modification and eventual merging of the plateaus upon increasing $\Delta$. 
}
    \label{plateaus}
\end{figure}

\subsection{Quantized SCO at $\omega_c\lesssim\Delta$}

First, we verify Eqs.~(\ref{33} and \ref{sco0}), indicating that at half-filling, we expect the $SCO/N_s$ to remain constant at very low magnetic fields, which corresponds to the zero magnetic field value in the absence of NNN hopping. We extract the cutoff $\alpha$ from this constant value using Eq.~\ref{sco0}, and demonstrate that it self-consistently aligns with the zero field estimate obtained from the Dirac spectrum.

Fig.~\ref{f2} depicts $SCO/N_s$ evaluated from the Hofstadter butterfly, showing the nonuniversal plateau at weak fields. The constant value is related to the cutoff $\alpha$, via Eq.~(\ref{sco0}). From here, one can extract the numerical value of the cutoff as:
\begin{equation}
\label{afa}
\alpha=0.464902\pm0.000047.
\end{equation}
This number is consistent with the UV energy scale corresponding to the linear dispersion of graphene within $7\%$ accuracy of the exact lattice dispersion. Fig. \ref{4} shows that the cutoff $\alpha$ is independent of $\Delta$ and system size $L$ for sufficiently large systems.

\begin{figure}[H]
    \centering
    \includegraphics[scale=0.25]{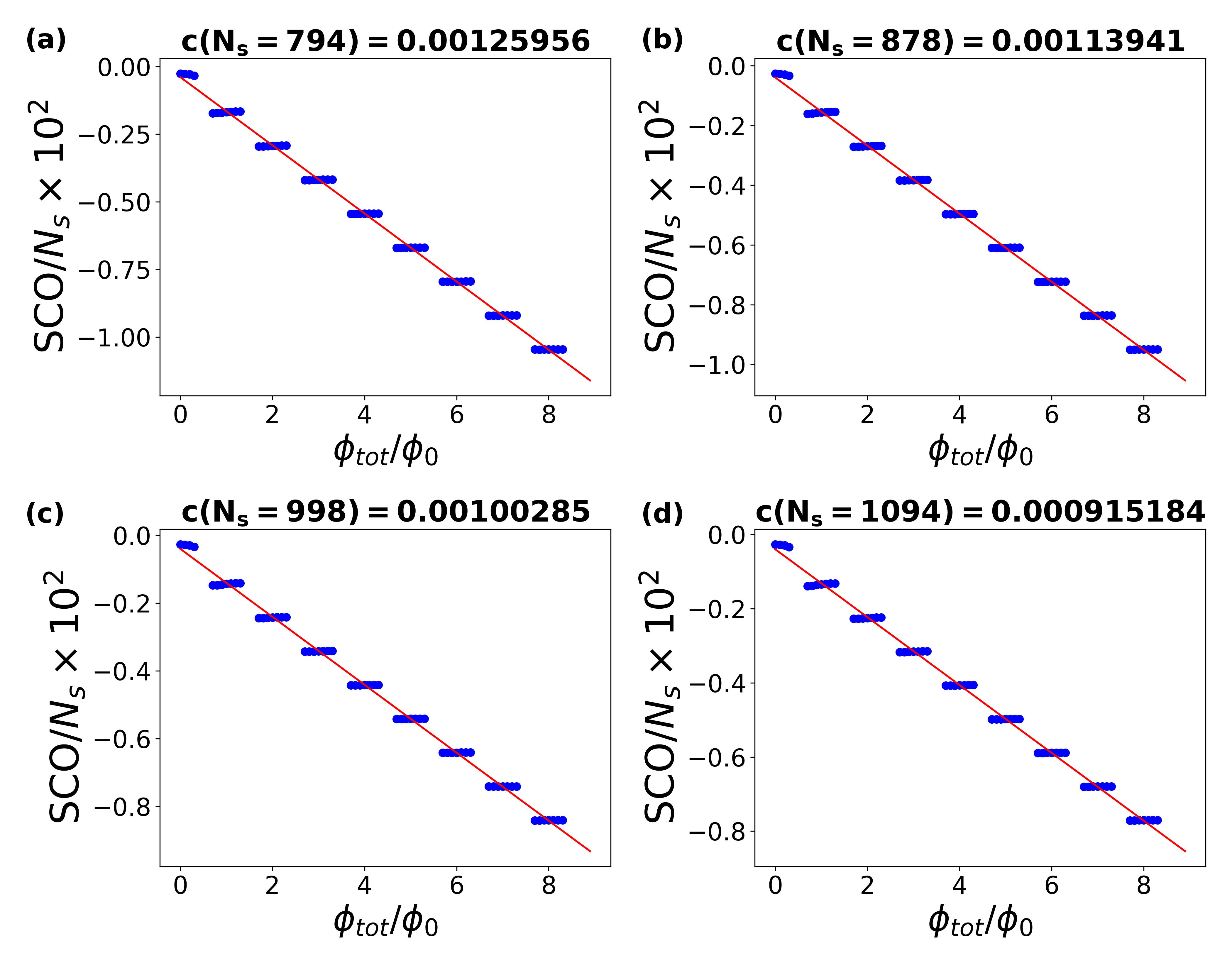}
    \caption{The SCO per site shows quantized plateaus at intermediate magnetic fields, with transitions between plateaus happening whenever the total magnetic flux through the system increases by an integer multiple of the flux quantum. The different panels show results for various system sizes. A linear fit to the plateau spacing across system sizes indicates that the gap between successive plateaus, $c(N_s)$, follows a universal quantization rule, being equal to $1/N_s$, where $N_s$ represents the number of sites. See subsequent Fig.~\ref{fig:c_vs_delta} for quantitative data. }
    \label{fig:linear fit}
\end{figure}

Next, we will focus on the behavior in intermediate fields, where $\omega_c\lesssim\Delta$, and observe that the SCO is quantized (see panel b and d in figure \ref{3}) as predicted by single electron transfer between sublattices per flux quantum, resulting from the sublattice polarization of the zeroth Landau level. We will examine the gaps between successive plateaus and demonstrate that they match our result. We will limit ourselves to field strengths corresponding to $\phi_B/\phi_0=0.5$, because a flux of $\frac{\phi_B}{\phi_0}$ and $1-\frac{\phi_B}{\phi_0}$ results in the same phase change across a unit cell. We will first examine the weak-sublattice potential regime $\Delta\ll\alpha$.

For weak-sublattice potentials, $\Delta \ll \alpha$, at intermediate magnetic field strengths, $SCO$ shows quantized plateaus. To compare our analytical computations of the previous section with the exact diagonalization results, let us address the following questions:
 At which discrete flux values do we have plateau transitions, and what is the gap between the successive plateau transitions?
For the first question, we plot $\frac{SCO}{N_s}$ vs the degeneracy of LL, given by $\frac{\phi_{tot}}{\phi_0}=N_{uc} \frac{\phi_B}{\phi_0}$, where $\phi_{tot}$ is the total flux through the sample and $N_{uc}$ is the number of unit cells.

\begin{figure}[H]
    \centering
    \includegraphics[height=4.5cm, width=8.5cm]{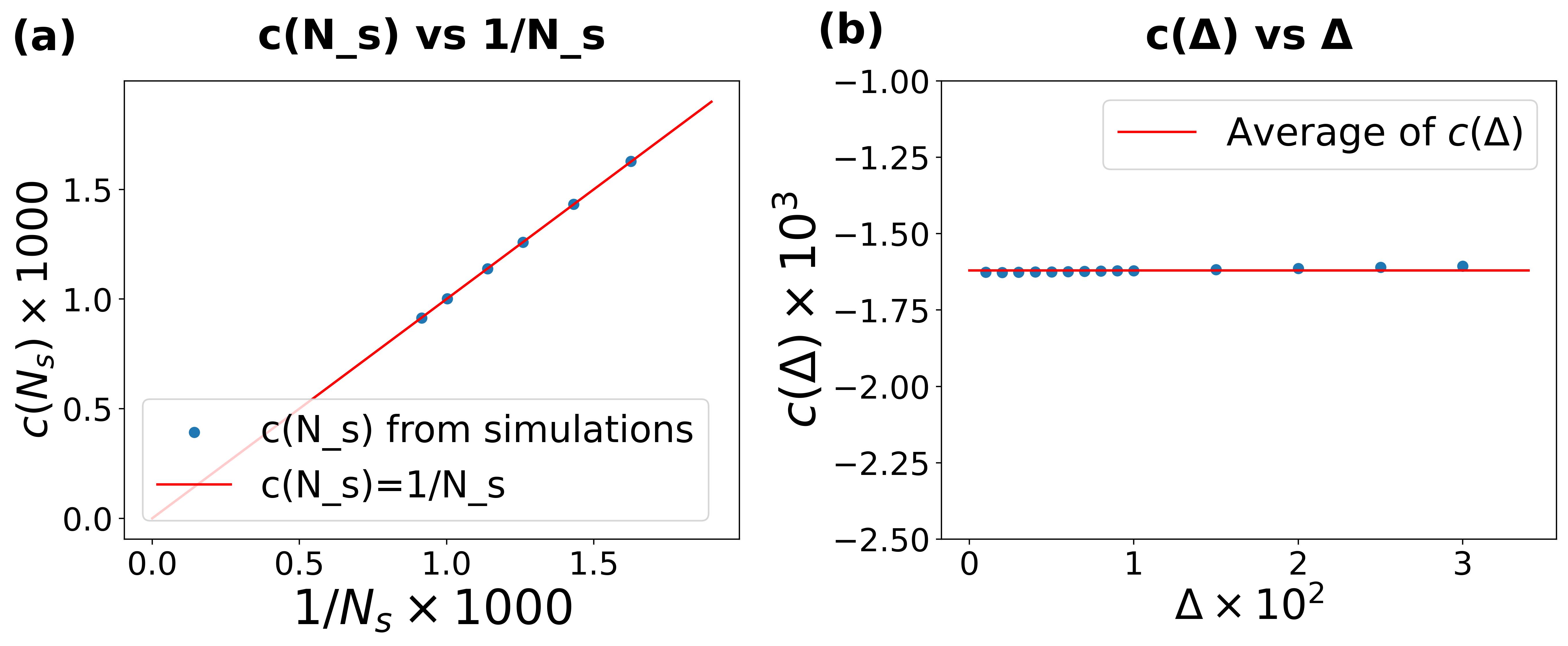}
    \caption{
    The gap $c$ between successive plateaus as a function of system size, $c(N_s)$ versus $1/N_s$ in panel (a), and as a function of sublattice potential, $c(\Delta)$ versus $\Delta$ in panel (b). The results demonstrate the parameter-independent universal quantization of SCO at intermediate fields. }
    \label{fig:c_vs_delta}
\end{figure}

We can therefore understand the plateaus in terms of the number of particles that occupy a LL, $d=[\frac{\phi_{tot}}{\phi_0}+0.5]$, where $[\ldots]$ denotes the smallest integer function. In terms of $d$, we observe that the jumps in the plateaus occur every time $d$ changes by an integer, i.e., when $\frac{\phi_{tot}}{\phi_0}=n+0.5$ ($n=1,2,3,...$). This is consistent with our analytical result, which demonstrated that the zeroth LL contributes $-1/N_s$ per particle to the $SCO/N_s$. Therefore, if the degeneracy of an LL changes by $1$, i.e., one more particle occupies the zeroth LL, we expect $SCO/N_s$ to jump by $-1/N_s$, according to Eq.~(\ref{33}). Furthermore, we plot $SCO/N_s$ against the total flux for different system sizes; the plateaus still occur at integer values of $d$ (see the panel (c) in Fig.~\ref{plateaus}).

From the analytical result in Eq.~\ref{33}, the spacing between successive plateaus is universal: it does not depend on the sublattice potential $\Delta$ and equals $N_s^{-1}$. Exact diagonalization confirms this prediction: a linear fit of $SCO$ plateau values across varying $\Delta$ and $N_s$ yields a slope consistent with the theoretical gap (see Fig.~\ref{fig:c_vs_delta}).
Moreover, we confirm that the gap is independent of the sublattice potential $\Delta$, as shown in panel (b) of Fig.~\ref{fig:c_vs_delta}. 

To summarize the discussion above, we have demonstrated, and verified through explicit lattice calculations, that the Dirac spectrum gives rise to quantized SCO at intermediate magnetic fields. This quantization is a direct consequence of the Dirac spectrum and, therefore, is a universal feature across all Dirac materials. However, the validity of the Dirac description is restricted: once the sublattice potential $\Delta$ is increased beyond the UV cutoff $\alpha$, the Dirac spectrum effectively ceases to exist.

\begin{figure}[H]
    \centering
    \includegraphics[height=5cm, width=8.5cm]{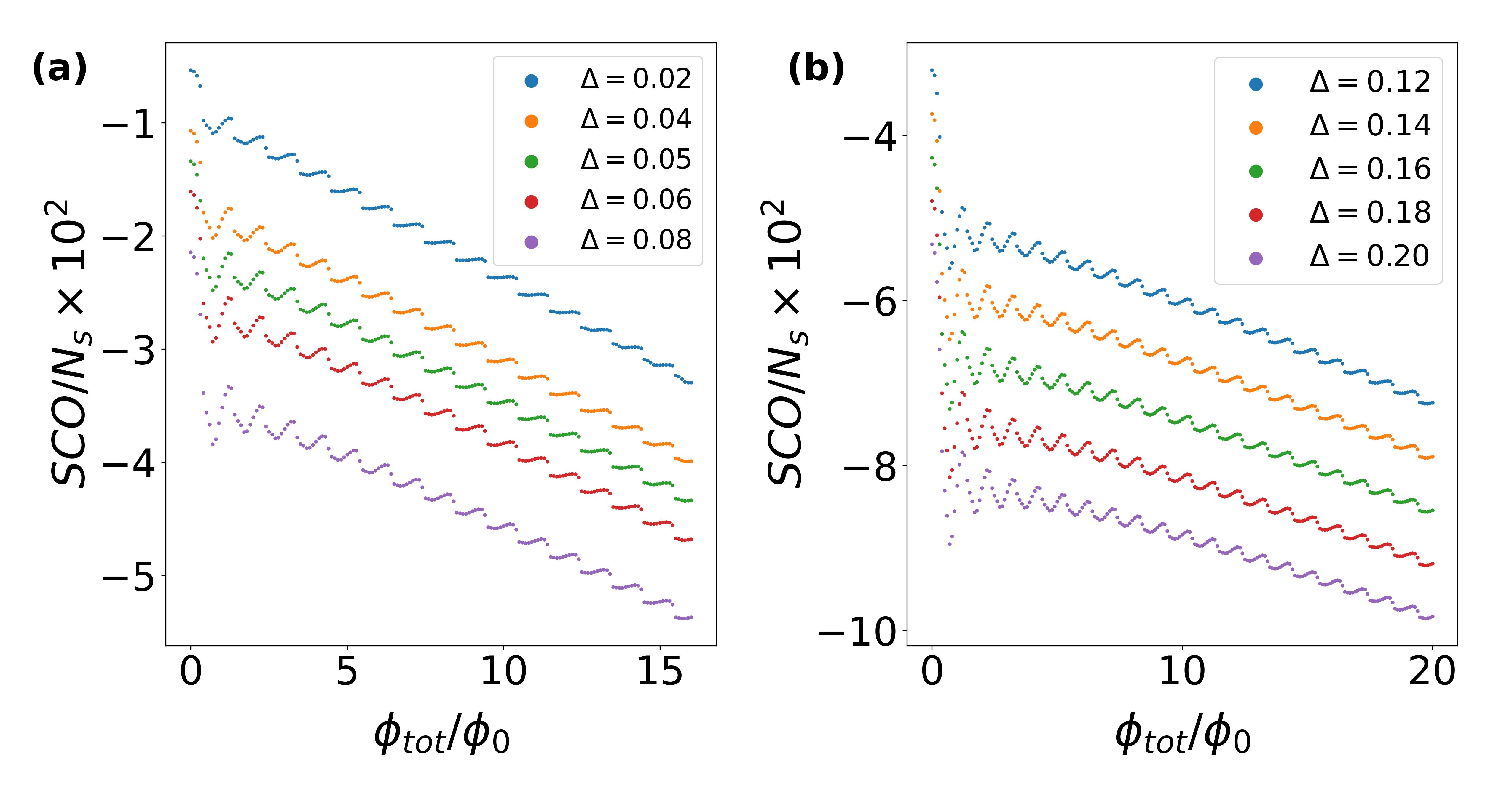}

   \caption{ Breaking of the universal quantization of the SCO outside the $\Delta\ll\alpha$ regime. As the sublattice potential $\Delta$ substantially increases, the quantized plateaus of the SCO gradually lose their robustness. In the opposite limit, $\Delta \gg \alpha$, the ultraviolet cutoff $\alpha$ associated with the Dirac dispersion no longer determines the low-energy physics. As a result, the effective Dirac description breaks down, and the universal quantization of the SCO characteristic of Dirac quasiparticles is lost.}

   \label{oscillations instead of plateau}
\end{figure}

Motivated by this observation, we systematically investigated the dependence of SCO on the sublattice potential. When $\Delta$ grows to values comparable to or larger than the cutoff $\alpha$, SCO is no longer quantized. This breakdown occurs because the Dirac approximation is invalid at these energy scales, and the behavior of the system can only be accurately captured using full lattice calculations, e.g., via exact diagonalization.
As illustrated in Fig.~\ref{oscillations instead of plateau}, increasing $\Delta$ relative to $\alpha$ progressively distorts the quantized plateaus: the initially sharp steps become oscillatory and lose their robustness. Upon further increasing $\Delta$ beyond the cutoff scale, the plateau structure disappears entirely, and SCO evolves into a smooth, featureless curve, as shown in panel (d) of Fig.~\ref{8}.

\subsection{Quantum oscillations at $\omega_c\gtrsim \Delta$ }
The behavior under strong magnetic fields appears qualitatively similar for both weak and strong sublattice potentials, exhibiting pronounced magneto-oscillations akin to the anomalous De Haas-Van Alphen oscillations observed in insulators at low temperatures, featuring dips at integer filling of LLs. 

\begin{figure}[H]
    \centering    \includegraphics[height=8cm, width=8.5cm]{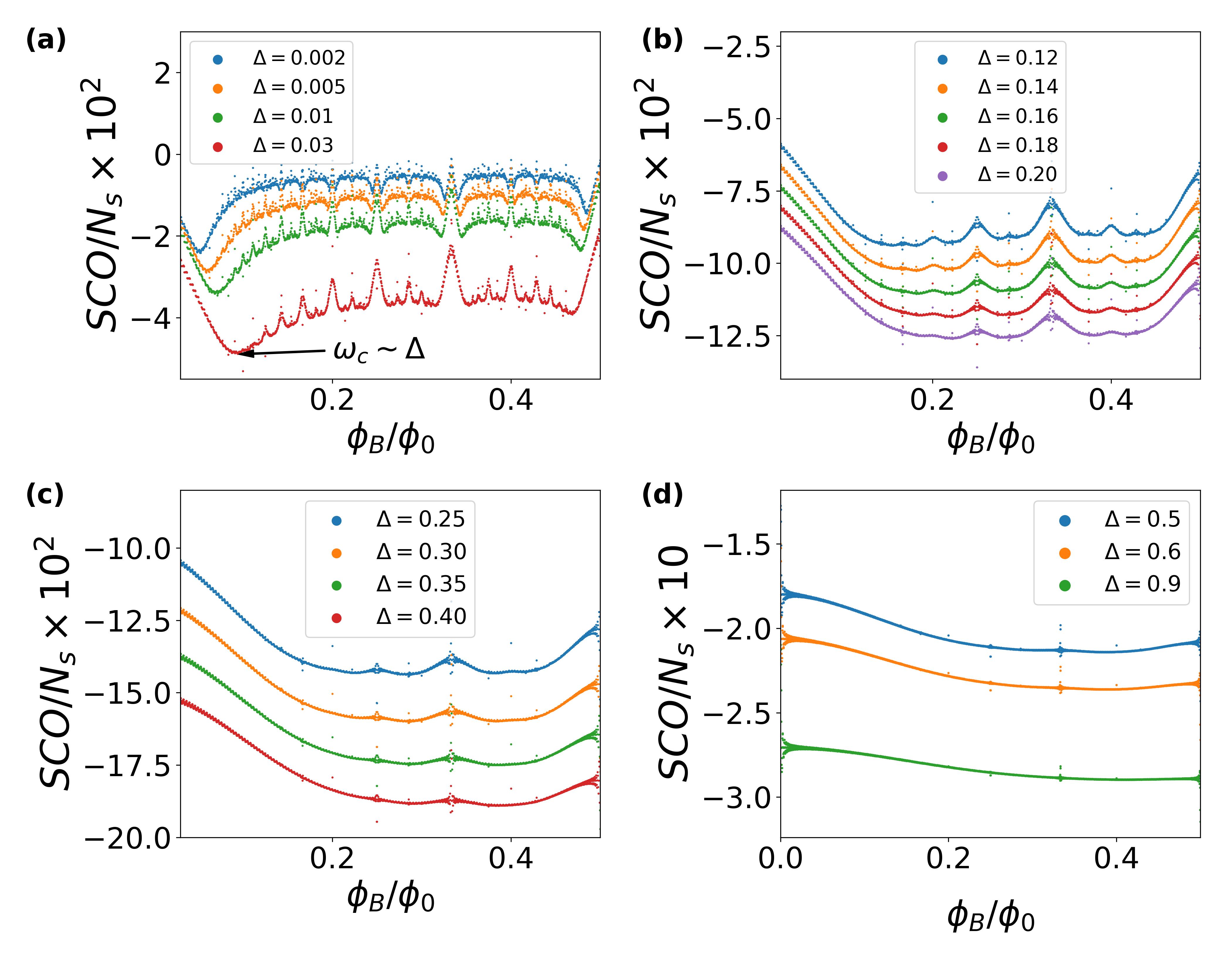}
    
    \caption{SCO per site is shown as a function of the total magnetic flux threading the system, measured in units of the flux quantum. The results span magnetic fields from the intermediate regime - where SCO exhibits quantized plateaus - to the strong field regime characterized by SCO magneto-oscillations. Varying the sublattice potential $\Delta$ reveals that the amplitude of these magneto-oscillations diminishes as $\Delta$ increases, reflecting a progressive departure from the regime $\Delta \ll \alpha$.}
    
    \label{8}
\end{figure}

The field strength at which these oscillations begin also increases with $\Delta$ as dictated by the condition $\omega_c \simeq\Delta$. The magneto-oscillations are stronger at weaker sublattice potentials, that is, when $\Delta\ll\alpha$. The magneto-oscillations diminish in strength as the sublattice potential increases. This is illustrated in Fig.~\ref{8}.  The number of filled LLs at a given magnetic field is $\phi_0/\phi_B$. As discussed above, the onset of these magneto-oscillations can be explained similarly to the onset of the anomalous de Haas-Van Alphen effect in insulators \cite{Qos}. They occur when the cyclotron frequency satisfies $\omega_c \gtrsim \Delta$ which explains the delay in their onset as we increase $\Delta$, as observed in Fig.~\ref{8}. These quantum oscillations are directly associated with the Dirac spectrum, consistent with the analysis of Ref.~\cite{Qos}, where a similar form of dispersion was considered. The connection is further supported by the progressive suppression and eventual disappearance of the oscillations upon increasing the sublattice potential. In the limit where $\Delta \gg \alpha$, the Dirac spectrum ceases to set the dominant energy scale, and correspondingly, the oscillatory behavior vanishes.

\begin{figure}[H]
    \centering
    \includegraphics[scale=0.5]{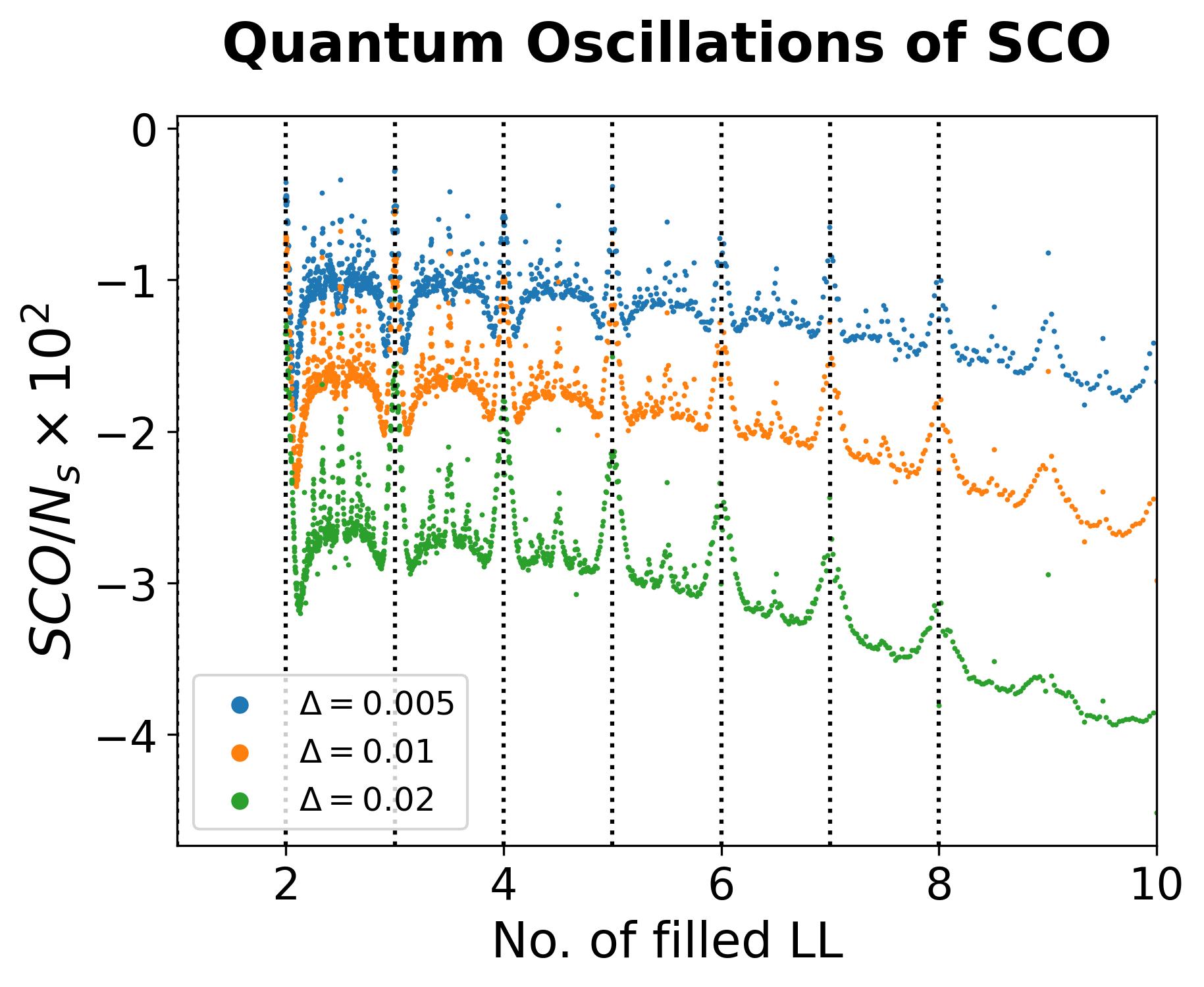}
    \caption{Quantum oscillations of SCO: SCO per site is shown as a function of Landau-level filling at strong magnetic fields. Clear oscillations are observed, with distinct dips occurring exactly at integer fillings, indicating full occupation of individual Landau levels.}
    \label{llsfill}
\end{figure}

A natural question arises regarding the behavior of the SCO in the regime $\omega_c>\alpha>\Delta$, where the cyclotron energy exceeds both the UV scale $\alpha$ and the zero-field spectral gap $\Delta$. Although in this regime $\alpha$ still defines the overall energy scale that determines the universal properties of the system, the dominant physical scale could now be the magnetic field through the cyclotron gap, $\omega_c$. One may therefore ask whether strong magnetic fields, by enhancing lattice effects, could suppress the SCO magneto-oscillations that originate from the Dirac nature of the zero-field low-energy spectrum. 

An inspection of the Hofstadter butterfly spectrum, Fig.~\ref{f1}, provides a clear picture. At large magnetic flux per plaquette, the higher LLs split into subbands and progressively lose their degeneracy due to the underlying lattice periodicity. In contrast, the lowest LLs retain their Dirac-like character and remain nearly dispersionless (degenerate for which the degeneracy is not lifted, i.e., unbranching), preserving the universal features of relativistic Landau quantization. Although the portion of the Dirac sea exhibiting universal Dirac behavior shrinks as the magnetic field increases, the degeneracy of the lowest unbranching LLs grows proportional to the magnetic flux. Consequently, a larger fraction of electrons continues to occupy these relativistic lowest LL states.
This enhanced occupation compensates for the shrinking universal region in the spectrum, ensuring that the SCO continues to exhibit pronounced magneto-oscillations tied to the Dirac low-energy structure. Detailed analysis of the Hofstadter butterfly confirms this conclusion: despite the full inclusion of lattice-induced phase effects accumulated by electrons traversing the magnetic unit cell, no qualitative change is observed in the oscillatory behavior of SCO across the $\omega_c>\alpha>\Delta$ regime. Hence, the universal Dirac contribution to magneto-oscillations remains robust even deep in the strong-field limit, as seen in Figs.~\ref{8}a and \ref{llsfill}.

\begin{figure}[H]
    \centering
    \includegraphics[height=5cm, width=8.5cm]{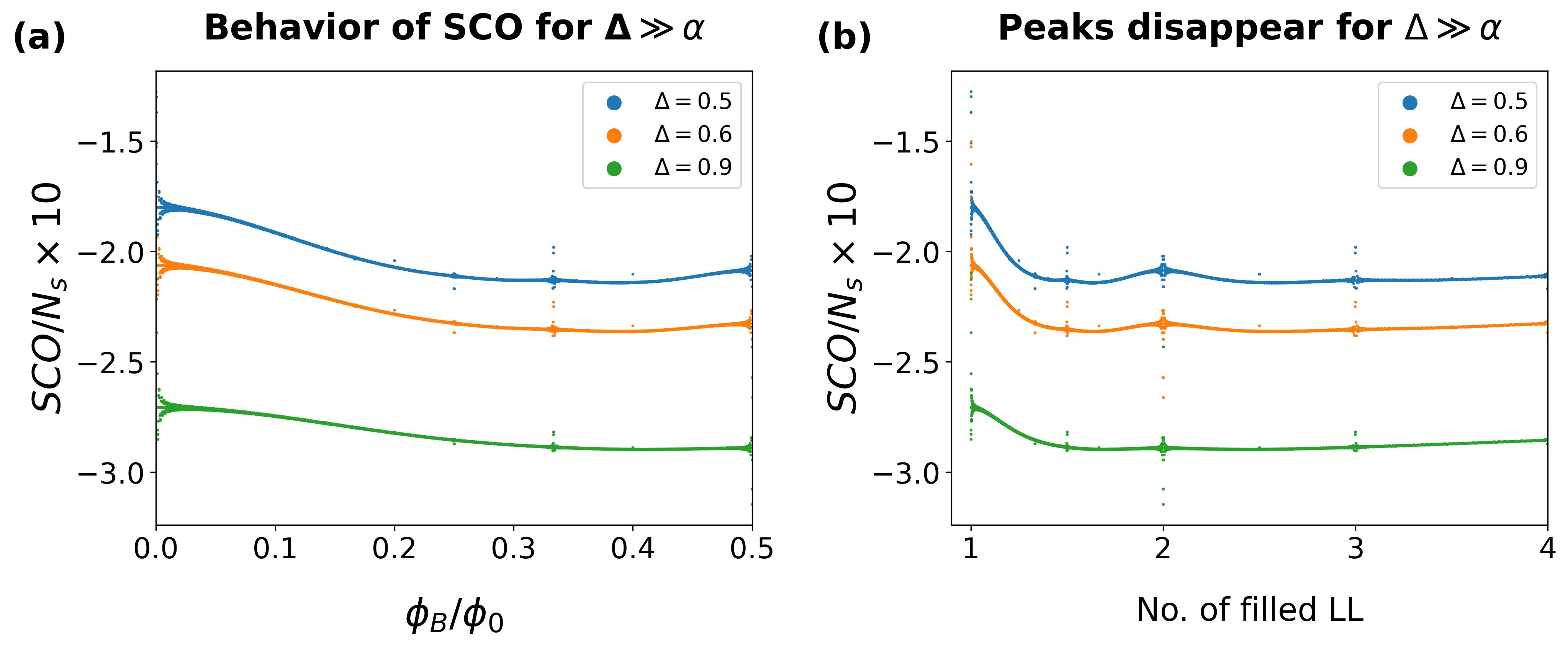}
    
    \caption{Disappearance of magneto-oscillations for $\Delta \gg \alpha$. The SCO magneto-oscillations are being suppressed upon increasing $\Delta$. For $\Delta \gg \alpha$, the ultraviolet cutoff $\alpha$ of the Dirac dispersion no longer sets the dominant energy scale. As a result, the effective Dirac description ceases to be valid, and the characteristic magneto-oscillatory behavior associated with Dirac quasiparticles disappears.}
    \label{dips lessen}
\end{figure}

To support our prior discussion and gain further insight into the nature of universal quantum magneto-oscillations, we draw the reader's attention to the behavior of $SCO$ versus the number of filled LLs, demonstrating that the dips in the magneto-oscillations of SCO occur exactly when an LL is filled. This is demonstrated in Fig.~\ref{llsfill}. In this regime, one should also realize that at strong quantizing magnetic fields, the typical number of filled LLs is relatively small ($\lesssim10$), as can be seen in Fig.~\ref{dips lessen}(b). When the sublattice potential exceeds $\alpha$, $\Delta > \alpha$, the ultraviolet cutoff $\alpha$ associated with the Dirac dispersion no longer controls the low-energy behavior. In this regime, the effective Dirac description is no longer valid, and as a result, the quantum oscillations of SCO, which are associated with Dirac quasiparticles, are vanishing leading to a continuous curve for SCO.

\section{Results and Discussion}

We demonstrate that Dirac insulators with broken sublattice symmetry exhibit a singular magnetic-field dependence of the SCO. The qualitative behavior of SCO is controlled by both the staggered sublattice potential, $\Delta$, and the magnetic field, with the relevant ultraviolet energy scale set by the cutoff $\alpha$—the range within which the lattice dispersion can be faithfully approximated by a Dirac Hamiltonian.

For Dirac insulators, namely weak sublattice potentials, $\Delta \ll \alpha$, SCO shows two distinct regimes. At intermediate magnetic fields, where $\omega_c \lesssim \Delta$, SCO exhibits universal quantization, originating from the integer degeneracy of the zeroth Landau level. At stronger magnetic fields, $\omega_c \gtrsim \Delta$, SCO develops magnetic oscillations analogous to the anomalous de Haas–van Alphen effect, as observed in Dirac insulators under strong fields at low temperature.
When the sublattice potential exceeds the cutoff, $\Delta \gtrsim \alpha$, the Dirac approximation no longer accurately describes the system. In this regime, the quantized behavior of SCO is lost, and its properties can only be captured through explicit lattice-level calculations, such as exact diagonalization.

As we discussed in the introduction, the predicted effects of quantization and magnetooscillations of SCO can be observed experimentally. In the intermediate-field regime, $\omega_c\lesssim\Delta$, for spinless (spin-polarized) electrons on a honeycomb lattice with SCO, the low-energy physics near the Dirac points is described by a massive Dirac Hamiltonian, yielding relativistic Landau levels $E_{n}^{\pm}/(-t)=\pm\sqrt{\Delta^2+2n\phi_b}$ ($n\ge1$) and an $n=0$ level at $E_0^{\pm}=\pm\Delta$ whose wave functions are polarized onto single sublattices. This sublattice polarization, together with the gap $2|\Delta|$, underlies the STM signatures of SCO in a magnetic field.
In this regime the sublattice-resolved local density of states (LDOS) satisfies $\rho_{A,B}(E)\approx\tfrac12\rho_{\rm tot}(E)\,[1\pm \Delta/E]$, so the directly measurable contrast $\Delta\rho(E)\equiv(\rho_A-\rho_B)/(\rho_A+\rho_B)\approx\Delta/E$ is largest near $|E|\simeq|\Delta|$ and reverses sign between occupied (negative-bias) and empty (positive-bias) states, consistent with the massive-Dirac description of graphene-like systems~\cite{electronicpropofgraphene,qhegraphene}. In scanning tunneling spectroscopy (STS), the differential conductance follows the Tersoff--Hamann\cite{TersoffHamannPRB1985} relation $dI/dV(\mathbf r,V)\propto\rho(\mathbf r,E=eV)$, so SCO can be detected and quantified by (i) atomically registering A vs B sites from a high-resolution topograph, (ii) acquiring grid STS (CITS) and binning the spectra by sublattice to obtain $\rho_A(E)$ and $\rho_B(E)$, (iii) forming a sublattice-odd map by adding $+dI/dV$ on A and $-dI/dV$ on B to image domains and domain walls, and (iv) tracking with $B$ the \emph{energy-integrated} sublattice-odd spectral weight over narrow windows around fully filled level $E_0^{-}$: when the magnetic flux increases by one flux quantum and the chemical potential lies in the appropriate gap, the integrated imbalance exhibits discrete steps fixed by the Landau-level degeneracy (per spin channel here), i.e., one electron transferred between sublattices per $\Phi_0$ - a universal quantization diagnostic accessible to STM ~\cite{TersoffHamannPRB1985,HoffmanScience2002,E.Andrei2009,MillerScience2009}. For reliable measurement two resolution conditions are required simultaneously: (a) \emph{Landau-level resolvability} - the relevant level spacing for $\omega_c<\Delta$ (e.g., the $0\!\leftrightarrow\!1$ spacing $\Delta E_{01}\approx\sqrt{\Delta^2+2\phi_b}-|\Delta| \sim \phi_b/|\Delta|$ in this limit) must exceed the total spectral broadening $\delta E\simeq\max\{3.5k_BT,\;1.8\,eV_{\rm rms},\;\Gamma\}$ set by temperature, lock-in modulation, and intrinsic lifetime/disorder, and (b) \emph{sublattice selectivity} - $|\Delta|$ itself should exceed $\delta E$ so that the A/B contrast near $E\approx\pm\Delta$ (and its characteristic sign reversal between occupied/empty states) remains strong~\cite{PhysRevB.50.4561,CrampinPRB2005,AstNatComm2016}. In practice, at $T\lesssim 4$ K with $V_{\rm rms}=0.3$--$0.8$ mV and clean samples ($\Gamma\sim1$--$3$ meV), a \emph{reasonable SCO energy scale} is $|\Delta|\gtrsim 5$ - $20$ meV (comfortably above $\delta E$ and still compatible with $\omega_c<\Delta$ in moderate fields), while a practical target for the intermediate-regime spacing is $\Delta E_{01}\gtrsim 3$ - $5$ meV to resolve flat, field-tunable plateaus in the integrated sublattice-odd signal (that changes sign under the interchange $A\leftrightarrow$B). Graphene on hBN (hexagonal boron nitride) provides a representative platform where sublattice symmetry breaking yields meV to tens of meV gaps and clear Landau quantization in STS, making it well suited for observing the universal quantization of SCO by the protocol above~\cite{HuntScience2013,electronicpropofgraphene,RevModPhys.83.1193,E.Andrei2009,MillerScience2009}.

Similarly, at strong fields, STM can resolve magnetooscillations of the SCO at sufficiently low temperature and in sufficiently clean samples, because as LLs sweep through the chemical potential, the sublattice-odd LDOS modulates with magnetic field and produces oscillation-like changes in sublattice-binned $dI/dV$ signals, enabling clear SCO magnetooscillations in STM. The same LL physics implies Shubnikov-de Haas magnetooscillations in transport as a function of $1/B$. Importantly, the crossover value $\omega_c=\Delta$ corresponds to the strongest A/B density imbalance (a pronounced extremum of SCO vs.\ field). In the vicinity of this transition regime, the sublattice contrast is optimal for STM measurements, enabling the precise observation of universal SCO quantization.

The obtained results have implications for interaction effects in doped Dirac semiconductors under a weak magnetic field that are known to produce nonanalytic corrections \cite{Wang2021Persistent,Wang2021InteractionGr,Wang2022Ballistic} based on the effect of the curving of quasiclassical trajectories \cite{Sedrakyan2008Interaction2DEG,Sedrakyan2008Crossover,Sedrakyan2008Magneto,Sedrakyan2007ZeroBias,Sedrakyan2007Smearing} and due to the magnetic phase of the pseudospin due to Dirac dispersion \cite{Wang2021Persistent}. It would be interesting to introduce SCO and investigate how its field dependence modifies these observables both at half-filling and away from charge neutrality. Investigating the effect of interlayer interactions and spin-orbit couplings in bilayer heterostructures in an external magnetic field is another interesting venue of applicability of the derived theory.

Another interesting open problem is the following: We have considered the staggered potential due to the strain to be constant. However, one can consider its modification in the presence of a $U(1)$ gauge field microscopically originating from the strain-induced staggered potential as discussed in \cite{PhysRevB.103.L201104}. However, modifications due to that would be subleading, with the leading correction being the one reported in this paper.

A different avenue for potential impact of the developed theory is in 2D quantum antiferromagnetism. It is established that the Chern-Simons superconductivity \cite{Sedrakyan2017Topological,Wang2022Chern,Sedrakyan2020Helical,Wang2018Antiferromagnetic,PhysRevB.106.L121117} of fermionized quantum XY antiferromagnets in two dimensions describes the Néel-ordered phase through the pairing of Chern–Simons fermions that exhibit a Dirac-like dispersion. When an additional Ising interaction is introduced in XXZ antiferromagnets, one expects the emergence of an out-of-plane Ising order, which corresponds directly to an SCO for Chern–Simons fermions. Given the results of the present work, it would be interesting to include the dependence of SCO on the underlying configuration of the Chern–Simons gauge field. 
It will be interesting to see how this field-dependent SCO modifies the self-consistent Chern–Simons gauge theory and, in turn, the spectrum of the Chern–Simons superconducting state for XXZ magnets. In particular, investigating the interplay between the SCO and the dynamical Chern–Simons flux could reveal novel gap structures or topological transitions within the superconducting phase.

On a different front, we have demonstrated that $SCO$ is modified by Maxwell-type flux in Dirac materials exhibiting broken sublattice symmetry. At this stage, it is natural to inquire about the effect of Chern-Simons type flux on $SCO$, for instance, in Haldane-Chern insulators \cite{haldane,PhysRevB.100.125428} within an external magnetic field. In this paper, we have not addressed the valley degeneracy. It is also interesting to consider the implications of breaking valley symmetry. A suitable platform to investigate both effects would be Haldane's Chern insulator with a flux distribution within the unit cell, which can be realized in spin-doped graphene \cite{Hill_2011}.

Finally, an interesting implication of the present work is in the field of frustrated XY antiferromagnets on a moat-band lattice that stabilize a chiral spin liquid (CSL) coexisting with out-of-plane Ising antiferromagnetic order \cite{Sedrakyan2014Absence,Sedrakyan2015Spontaneous,Sedrakyan2015Statistical,WeiSedrakyan2023,PhysRevLett.111.257201}. Upon Chern–Simons fermionization, this CSL state is equivalent to a system of Chern–Simons Dirac fermions with SCO, which is equivalent to the original Ising spin ordering filling a topologically nontrivial band with Chern number 1. These Dirac fermions are minimally coupled to a U(1) gauge field governed by a statistical Chern–Simons action. Given that our analysis suggests the SCO amplitude acquires an explicit dependence on the Chern–Simons flux, it is natural to explore how this field dependence influences the low-energy effective theory of the CSL. In particular, one should investigate whether the Chern–Simons field–dependent SCO can drive an unconventional quantum phase transition from the CSL to the Néel-ordered state.

 \section*{Acknowledgment}
 The work was supported by the Armenian ARPI Remote Laboratory program 24RL-1C024.

\section*{Data Availability}

All the figures present in the paper are openly available \cite{OSFfigures}.

\appendix

\section{SCO in absence of magnetic field}\label{A}

The energy eigenvalues for Hamiltonian \ref{13} are given by:
\begin{equation}\label{14}
    \frac{E_{\bar {\mathbf k}}^\pm}{-t}=r \bar{\mathbf k}^2\pm\sqrt{{\bar{\mathbf k}^2}+\Delta^2},
\end{equation}
and the wavefunction at $K$ valley is:
\begin{equation}
    \begin{split}
        \Psi_K^\mathbf k(\mathbf x)=&\frac{1}{\sqrt {2A}}\bigg(1\mp \frac{\Delta}{\sqrt{\bar k^2+\Delta^2}}\bigg)^{1/2}\\&\times\begin{pmatrix}
            (-\bar k_x+\ii \bar k_y)/(\pm\sqrt{\bar{\mathbf k}^2+\Delta^2}-\Delta) \\ 1
        \end{pmatrix}e^{\ii \mathbf k\cdot \mathbf x}
    \end{split}
\end{equation}
The SCO, $\Lambda_\eta^\mathbf k$, corresponding to the state with an energy level labeled by $\mathbf k$ can be cast in the form:
\begin{equation}
    \Lambda_K^\mathbf k=\pm\frac{\Delta}{\sqrt{\bar{\mathbf k}^2+\Delta^2}}.
\end{equation}
The wavefunction for the Hamiltonian without the staggered potential is obtained from here by taking the $\Delta=0$ limit
\begin{equation}
    \Psi_K^\mathbf k(\mathbf x)=\frac{1}{\sqrt {2A}}\begin{pmatrix}
        (\mp\bar k_x\pm\ii \bar k_y)/\bar {\mathbf k}^2 \\ 1
    \end{pmatrix}e^{\ii \mathbf k\cdot \mathbf x},
\end{equation}
where $A$ is the area of the sample (we need this to ensure $\int d^2x\, \Psi^\dagger(x) \Psi (\mathbf x)=1$). One can immediately see that in this case ($\Delta=0$), $\Lambda_K^\mathbf k=0$ and SCO vanishes.

\section{Peierl's Substitution in presence of NNN hopping} \label{D}

In the presence of NNN hopping, the Bloch Hamiltonian is given by:
\begin{equation}
    H(\mathbf p)=-t \begin{pmatrix}
       r T_\mathbf p^A & T_\mathbf p\\ T_\mathbf p^* & r T_\mathbf p^B
    \end{pmatrix}.
\end{equation}
Here, $rt$ denotes the NNN hopping amplitude and 
\begin{equation}
    T_\mathbf p^{A/B}
=e^{\ii \mathbf p\cdot\mathbf a_1}+e^{\ii \mathbf p\cdot\mathbf a_2}+e^{\ii \mathbf p\cdot\mathbf a_3}+c.c.,
\end{equation}
where $\mathbf a_1=a(1,0)$, $\mathbf a_2=a(\frac{1}{2},\frac{\sqrt 3}{2})$, and $\mathbf a_3=a(\frac{1}{2},-\frac{\sqrt 3}{2})$ (are the lattice translation vectors).
One can now observe that $T_\mathbf p^A=T_\mathbf p T_\mathbf p^*-3$ and $T_\mathbf p^A=T_\mathbf p^* T_\mathbf p-3$. However, once one performs the momentum elongation in the presence of a Maxwell-type magnetic field, then $T_\mathbf p^A\neq T_\mathbf p T_\mathbf p^*-3$ due to \ref{28}. One is thus left with the choice of using $T_\mathbf p^A$ or $T_\mathbf p T_\mathbf p^*$. It turns out that the correct choice is $T_\mathbf p^{A/B}$ since it is the natural option, because in the presence of a magnetic field, the NNN hopping is not equivalent to two NN hops due to the magnetic flux (as evident from Eq.~\ref{28}). Next, we diagonalize the Hamiltonian near the high symmetry points $K$ and $K^\prime$. 
Near the $K$ point, the Hamiltonian is given by:
\begin{equation}
    \mathcal H(K) =-t \begin{pmatrix}
        r\bar k^2 & -\bar k_x+\ii \bar k_y \\
        -\bar k_x-\ii \bar k_y & r\bar k^2
    \end{pmatrix}.
\end{equation}

In terms of the creation and annihilation operators, defined in Eq.~(\ref{5}), $\mathcal H(K)$this becomes:
\begin{equation}\label{28}
    \mathcal H(K) =-t \begin{pmatrix}
        2r\phi_b(a^\dagger a+\frac{1}{2}) & \sqrt{2\phi_b} \hat a \\ \sqrt{2 \phi_b} \hat a^\dagger & 2r\phi_b(a^\dagger a+\frac{1}{2})
    \end{pmatrix}.
\end{equation}
The eigenenergies and eigenfunctions of $\mathcal H(K)$ are given by:
\begin{equation}
    \begin{split}
        E_n^\pm&=2 n r\phi_b\pm\sqrt{2n\phi_b+r^2\phi_b^2}\\
        \Psi_K^{n,k}(\mathbf r)&\sim\begin{pmatrix}
            \Phi_{n-1,k}(\mathbf r)/(c_n\pm\sqrt{1+c_n^2})\\
            \Phi_{n,k}(\mathbf r)
        \end{pmatrix},
    \end{split}
\end{equation}
where $c_n=r\sqrt{\phi_b/2n}$. Notice that we haven't used an equality sign for the wavefunction because the answer is off by a normalisation factor (denoted by $\mathcal N_n^K$), which ensures $\int d^2\mathbf r\,\Psi^\dagger (\mathbf r)\Psi(\mathbf r)=1$.

\section{Beyond half-filling in absence of magnetic field}\label{C}

Upon introducing the Fermi-energy, $\epsilon_F=-\varepsilon_F t$, with corresponding Fermi momentum $k_F$ such that $\sqrt{k_F^2+\Delta^2}$ at zero temperature, the SCO will be given by:
\begin{equation}
    \Lambda_{\nu>\frac{1}{2}}=\int_{k_F}^{k_c} \frac{-2\Delta}{\sqrt{\bar{\mathbf k}^2+\Delta^2}}\frac{A}{2\pi}k\, dk,
\end{equation}
which yields the SCO beyond half filling to be:
\begin{equation}
    \frac{\Lambda_{\nu>\frac{1}{2}}}{N_s}=-\frac{\Delta}{\sqrt 3\pi}(\alpha-\varepsilon_F).
\end{equation}

One can relate the Fermi energy to the density of particles with the Dirac dispersion:
\begin{equation}                    \varepsilon_F^2=3\pi \rho a^2+\Delta^2.
\end{equation}
Then, at high densities, the SCO above half-filling is given by:
\begin{equation}\label{26} 
    \begin{split}
        \frac{\Lambda_{\nu>\frac{1}{2}}}{N_s}=&-\frac{\Delta}{\sqrt 3 \pi}\bigg[\alpha-\sqrt{3\pi\rho a^2}\bigg(1+\frac{\Delta^2}{6\pi\rho a^2}\bigg)\Bigg]\\&+O\bigg(\bigg(\frac{\Delta^2}{6\pi\rho a^2}\bigg)^2 \bigg)
    \end{split}
\end{equation}
And at low densities, the SCO is:
\begin{equation}\label{27}
    \begin{split}
        \frac{\Lambda_{\nu>\frac{1}{2}}}{N_s}=&-\frac{\Delta}{\sqrt 3 \pi}\bigg[\alpha-\Delta\bigg(1+\frac{3\pi\rho a^2}{2\Delta^2}\bigg)\bigg]\\&+O\bigg(\bigg(\frac{3\pi\rho a^2}{\Delta^2}\bigg)^2\bigg)
    \end{split}
\end{equation}.

One must notice that there is no dependence on the NNN hopping amplitude, meaning SCO is the same irrespective of whether NNN hopping is present or not. This is because a NNN hop in the absence of a magnetic field is equivalent to two subsequent NN hops; thus, adding the NNN term in the Hamiltonian does not affect either the wavefunction or the density of states, and hence the SCO is unchanged. However, in the presence of a magnetic field, a NNN hop is not equivalent to two NN hops because of the magnetic flux:
\begin{equation}
    e^{-\ii \mathbf p\cdot \mathbf a_3}e^{\ii \mathbf p\cdot \mathbf e_1}e^{-\ii \mathbf p\cdot \mathbf e_2}=e^{-\ii 2\pi\phi_B/\phi_0},
\end{equation}
where $\phi_B=B a^2/4\sqrt 3$ (flux through the triangle enclosed by two successive nearest neighbor hops from $A\xrightarrow{}B\xrightarrow{}A$ and the corresponding next nearest hop from $A\xrightarrow{} A$) and $\phi_0=2\pi/e$ (flux quantum in natural units, $\hbar =1$). This will lead to a correction that is $\sim r \Delta B$.

\section{SCO in presence of magnetic field}\label{B}

In the presence of a magnetic field, one can expand the Hamiltonian around a highnsymmetry point ($\mathbf k_0$) in the Brillouin zone and make use of momentum elongation: $\mathbf k (=\mathbf p-\mathbf k_0)\xrightarrow{} -\ii \mathbf \nabla+e \mathbf A\equiv \mathbf k$.\\

Near the $K$ point, $\mathbf k_0=\frac{1}{a}(\frac{4\pi}{3},0)$, the Bloch Hamiltonian after the elongation is given by:
\begin{equation}
    \mathcal H(K) =-t \begin{pmatrix}
        0 & -\bar k_x+\ii \bar k_y \\
        -\bar k_x-\ii \bar k_y & 0
    \end{pmatrix},
\end{equation}
where $\bar k_i\equiv \sqrt 3 a k_i/2$. The commutation relation between the covariant momenta $\bar p_i$ is given by:
\begin{equation}\label{ppcom}
    [\bar k_i,\bar k_j]=-i \epsilon_{ij}\phi_b,
\end{equation}
where $\phi_b=\sqrt{3}e \phi_B/2$, is dimensionless and $\phi_B$ is flux through an unit cell. These commutation relations have significant consequences for the electronic structure of the system; primarily, they lead to the quantization of the Landau levels with degenerate energy levels. To diagonalize the Hamiltonian, we can define the creation and annihilation operators using the commutation relation \ref{ppcom}:
\begin{equation}\label{5}
    \begin{split}
        \hat a&=\frac{1}{\sqrt{2\phi_b}}(-\bar k_x+\ii\bar k_y)\\
        \hat a^\dagger&=\frac{1}{\sqrt{2\phi_b}}(-\bar k_x-\ii\bar k_y)\\
        [\hat a,& \hat a^\dagger]=1
    \end{split}
\end{equation}

In terms of the creation and annihilation operators, the Bloch Hamiltonian at $K$ point is given by:
\begin{equation}\label{HK}
    \mathcal H(K)=-t \begin{pmatrix}
        0 & \sqrt{2\phi_b} \hat a \\ \sqrt{2 \phi_b} \hat a^\dagger & 0
    \end{pmatrix}
\end{equation}
This effective low energy Hamiltonian can be exactly solved. The energy is given by:
\begin{equation}
    \frac{E_n^{\pm}}{-t}=\pm\sqrt{2n\phi_b},
\end{equation}
and the wavefunction is given by:
\begin{equation}
    \Psi_K^{n,k}(\mathbf r)=\begin{pmatrix}
        \pm\Phi_{n-1,k}(\mathbf r)\\
        \Phi_{n,k}(\mathbf r)
    \end{pmatrix}.
\end{equation}
Here, $\Phi_{n,k}(\mathbf r)$ is defined similarly to harmonic oscillator:
\begin{equation}
    \begin{split}
        \Phi_{n,k}(\mathbf r)&=\frac{1}{\sqrt{n!}}(\hat a^\dagger)^n\Phi_0(\mathbf r)\\
        \hat a\Phi_0(\mathbf r)&=0
    \end{split}.
\end{equation}
The exact expression for $\Phi_{n,k}(\mathbf r)$ can be found from solving the differential equation:
\begin{equation}
    \hat a^\dagger \hat a\Phi_{n,k}(\mathbf r)=n\Phi_{n,k}(\mathbf r).
\end{equation}
The exact expression is given by:
\begin{equation}\label{37}
    \Phi_{n,k}(\mathbf r)=\frac{1}{\sqrt{2^n n! \sqrt \pi}} e^{\ii k x}e^{\frac{-\bar y^2}{2}}H_n(\bar y),
\end{equation}
where $\bar y\equiv( y-k l_b^2)/l_b$, $l_b^2=1/eB$ and $k$  corresponds to the quantum number for translations along $x$ (the degeneracy of the quantum level) and $n\in {\mathbf Z}_+$, which indicates the quantum level. It is important to note that this result holds in the Landau gauge. We will adhere to the Landau gauge for the remainder of the text. First, we observe that the spectrum is symmetric around $E=0$ (which is the zeroth Landau level). This symmetry arises because this Hamiltonian possesses the following symmetry:
\begin{equation}\label{chiralsymmetry}
    \{\mathcal H(K),\sigma_z\}=0.
\end{equation}
It ensures that if $\psi$ is an eigenstate with energy $E$, then $\sigma_z \psi$ is also an eigenstate with energy $-E$.

Second, we observe that the density of particles is uneven at sites A and B:
\begin{equation}
    \begin{split}
        |\psi_{A,K}^{(n,k)}(\mathbf r)|^2&=|\Phi_{n-1,k}(\mathbf r)|^2\\
        |\psi_{B,K}^{(n,k)}(\mathbf r)|^2&=|\Phi_{n,k}(\mathbf r)|^2
    \end{split}
\end{equation}
The SCO density at the K point, is therefore given by:
\begin{equation}\label{CDOK}
    \Lambda_K^{n,k}=(\Psi_K)^\dagger\sigma_z\Psi_K=|\Phi_{n-1,k}(\mathbf r)|^2-|\Phi_{n,k}(\mathbf r)|^2,
\end{equation}
where $\mathbf r=(x,y)$, $\mathbf r_0=(0,k l_B^2)$. This physically gives the SCO density contribution from the $K$ valley around $\mathbf r_0$ due the Landau level state labeled by $(n,k)$.
Near the $K^\prime$ point ($k_0=(-\frac{4\pi}{3a},0)$), the Hamiltonian is related to the Hamiltonian at $K^\prime$ by $k_x\xrightarrow{}-k_x$ and $k_y\xrightarrow{}k_y$ (which is the parity transformation in 2D), and is given by:
\begin{equation}
    \mathcal H(K^\prime) =-t \begin{pmatrix}
        0 & \bar k_x+\ii \bar k_y \\
        \bar k_x-\ii \bar k_y & 0
    \end{pmatrix}.
\end{equation}
One can diagonalise this exactly by defining the creation and annihilation operators:
\begin{equation}\label{17}
    \begin{split}
        \hat a&=\frac{1}{\sqrt{2\phi_b}}(\bar k_x-\ii\bar k_y)\\
        \hat a^\dagger&=\frac{1}{\sqrt{2\phi_b}}(\bar k_x+\ii\bar k_y)\\
        [\hat a,& \hat a^\dagger]=1.
    \end{split}
\end{equation}
This transforms Hamiltonian leading to
\begin{equation}\label{HK'}
    \mathcal H(K^\prime)=-t \begin{pmatrix}
        0 & \sqrt{2\phi_b} \hat a^\dagger \\ \sqrt{2 \phi_b} \hat a & 0
    \end{pmatrix}.
\end{equation}
This Hamiltonian can be exactly solved and the energies and corresponding wavefunctions are given by:
\begin{align}
    \frac{E_n^{\pm}}{-t}&=\pm\sqrt{2n\phi_b}\\
    \Psi_{K^\prime}^{n,k}(\mathbf r)&=\frac{1}{\sqrt 2}\begin{pmatrix}
        \Phi_{n,k}(\mathbf r)\\
        \pm\Phi_{n-1,k}(\mathbf r)
    \end{pmatrix},
\end{align}
where $\Phi_{n,k}(\mathbf r)$ has the same form as the at $K$ point. Comparing the wavefunctions we can observe that they are related by:
\begin{equation}\label{1.1.21}
    \Psi_{K^\prime}^{n,k}=\sigma_x\Psi_K^{n,k}.
\end{equation}
This relation is due to fact that the Hamiltonian operators near the two valleys are related as follows (from equation \ref{HK} and \ref{HK'}):
\begin{equation}\label{1.1.22}
    \mathcal H(K^\prime)=\sigma_x\mathcal H(K)\sigma_x.
\end{equation}
This relation is reminiscent of the fact that the two valleys are connected by a parity transformation (unlike \ref{4,5}, they are no longer related by time-reversal due to the presence of a magnetic field) and the sublattice symmetry. Using equation \ref{1.1.21} and $\{\sigma_i,\sigma_j\}=2 \delta_{ij}$, one can immediately observe that:
\begin{equation}\label{1.1.23}
    \Lambda_K^{n,k}+\Lambda_{K^\prime}^{n,k}=0
\end{equation}

This means that the sublattice charge order is overall canceled out between two valleys related by parity transformation. Note that this symmetry argument extends beyond the linearized spectrum we have been working with so far. This has been verified numerically by observing a zero SCO at strong magnetic fields. In conclusion, there is no overall sublattice charge order in graphene with nearest-neighbor hopping in the presence of sublattice symmetry. The situation remains unchanged even if we include next-nearest neighbor hopping because it does not break the sublattice symmetry, and the sublattice charge order still cancels out between the valleys.

\section{Diagonalizing Dirac Hamiltonian in presence of sublattice potential} \label{E}

In the presence of the sublattice potential $\Delta$ and at $r=0$ 
the Hamiltonian near the $K$ point is given by:
\begin{equation}\label{22}
    \mathcal H(K)=-t \begin{pmatrix}
        \Delta & \sqrt{2\phi_b} \hat a \\ \sqrt{2 \phi_b} \hat a^\dagger & -\Delta
    \end{pmatrix}.
\end{equation}
Diagonalization of this Hamiltonian gives us the following wavefunctions and energies:
\begin{equation}
    \begin{split}    \Psi_K^{n,k}(\mathbf r)&=\mathcal N_n^K\begin{pmatrix}
            \Phi_{n-1,k}(\mathbf r)\\
            \left(-\Delta_n\pm\sqrt{1+\Delta_n^2}\right)\Phi_{n,k}(\mathbf r)
        \end{pmatrix},\\
        \frac{E_n^\pm}{-t}&=\pm\sqrt{2n\phi_b+\Delta^2},\\
         |\mathcal N_n^K|^2&=\frac{1}{1+(c_n^K)^2}.
    \end{split}
\end{equation}

Similarly, the wavefunctions at $K^\prime$ are given by (energies remain unchanged):
\begin{align}
\Psi_{K^\prime}^{n,k}(\mathbf r)&=\mathcal N_n^{K^\prime}\begin{pmatrix}
            (\Delta_n\pm\sqrt{1+\Delta_n^2})\Phi_{n,k}(\mathbf r)
            \\
            \Phi_{n-1,k}(\mathbf r)
        \end{pmatrix}\\
        |\mathcal N_n^{K^\prime}|^2&=\frac{1}{1+(c_n^{K^\prime})^2}
\end{align}
where $c_n^{K^\prime}=\Delta_n\pm\sqrt{1+\Delta_n^2}$. The total SCO density from the two valleys is:
\begin{equation}
\begin{split}  
    &\Lambda_{n,k}^K(\mathbf r-\mathbf r_0)+\Lambda_{n,k}^{K^\prime}(\mathbf r-\mathbf r_0)\\&=|\Phi_{n-1,k}|^2\bigg(\frac{|\mathcal N_n^K|^2}{(\Delta_n\mp\sqrt{1+\Delta_n^2})^2}-\frac{|\mathcal N_n^{K^\prime}|^2}{(\Delta_n\pm\sqrt{1+\Delta_n^2})^2}\bigg)\\
    &+|\Phi_{n,k}|^2\bigg(|\mathcal N_n^K|^2-|\mathcal N_n^{K^\prime}|^2\bigg)
\end{split}
\end{equation}
The $\pm$ sign corresponds to the SCO density associated with the energy level $E_n^\pm$. At half filling, only the LLs with negative energy contribute; thus, we mostly consider the lower sign in this article.

\section{Corrections to SCO in presence of NNN hopping}\label{NNN section}

In the presence of NNN hopping, the Bloch Hamiltonian of graphene with the staggered sublattice potential is given by:
\begin{equation}
    H(\mathbf p)=-t \begin{pmatrix}
       r T_\mathbf p^A+\Delta & T_\mathbf p\\ T_\mathbf p^* & r T_\mathbf p^B-\Delta
    \end{pmatrix}.
\end{equation}
Near the $K$ point, the Hamiltonian is given by:
\begin{equation}
    \mathcal H(K) =-t \begin{pmatrix}
        r\bar k^2+\Delta & -\bar k_x+\ii \bar k_y \\
        -\bar k_x-\ii \bar k_y & r\bar k^2-\Delta
    \end{pmatrix}.
\end{equation}
This Hamiltonian can also be diagonalized exactly. The energies and wavefunctions are given by:
\begin{equation}
    \begin{split}    \Psi_K^{n,k}(\mathbf r)&=\mathcal N_n^K\begin{pmatrix}
            \Phi_{n-1,k}(\mathbf r)\\
            \left(c_n-\Delta_n\pm\sqrt{1+(c_n-\Delta_n)^2}\right)\Phi_{n,k}(\mathbf r)
        \end{pmatrix},\\
        \frac{E_n^\pm}{-t}&=2nr\phi_b\pm\sqrt{2n\phi_b+\Delta^2+r^2\phi_b^2-2r\Delta\phi_b.}\\
    \end{split}
\end{equation}
where $\mathcal N_n^K$ is the normalization that can be easily computed. The Hamiltonian near $K^\prime$ point can again be obtained by replacing $k_x\xrightarrow{}-k_x$ and $k_y\xrightarrow{}k_y$. Diagonalization of this Hamiltonian yiels the wavefunction in the form:
\begin{equation}
    \Psi_{K^\prime}^{n,k}(\mathbf r)=\mathcal N_n^{K^\prime}\begin{pmatrix}
            \left(c_n+\Delta_n\pm\sqrt{1+(c_n+\Delta_n)^2}\right)\Phi_{n,k}(\mathbf r)
            \\
            \Phi_{n-1,k}(\mathbf r)
            
        \end{pmatrix},
\end{equation}
where $\mathcal N_n^{K^\prime}$ is, again, the normalization.
The contribution to the SCO from the two valleys from a LL is then given by:
\begin{align}\label{sconnn}
    \Lambda_{n,k} = 
    \bigg[ &
    4\Delta_n c_n 
    - (c_n + \Delta_n)\sqrt{1 + (c_n + \Delta_n)^2} \nonumber \\
    &+ (c_n - \Delta_n)\sqrt{1 + (c_n - \Delta_n)^2}
    \bigg]/\\\nonumber
    &\bigg[1+2(\Delta_n^2+c_n^2)+(c_n^2-\Delta_n^2)\bigg[c_n^2-\Delta_n^2\\\nonumber
    &+(c_n-\Delta_n)\sqrt{1+(c_n+\Delta_n)^2}\\\nonumber
    &+(c_n+\Delta_n)\sqrt{1+(c_n-\Delta_n)^2}\\\nonumber&+\sqrt{1+2(c_n^2+\Delta_n^2)+(c_n^2-\Delta_n^2)^2}\bigg]\bigg].
\end{align}

The zeroth LL is unaffected by the NNN hopping, and the wavefunction is still given by: $\Psi^{K}_{0,k}(\mathbf r)=\left(0\,\,\,\Phi_{n,k}(\mathbf r)\right)$; and hence the SCO contribution is given by $-1$. The energy of the zeroth LL at $K$ valley is given by $-\Delta+r\phi_b$. At weak sublattice potential $r\Delta\ll1$ and hence, $\Delta^2\ll\Delta/r$; which implies that in the quantized SCO regime, $\phi_b\ll\Delta^2$, $r \phi_b\ll\Delta$. Thus, even in the presence of NNN hopping, only the $K$ valley contributes to the half filling when $\omega_c\ll\Delta$. Furthermore, when $r\phi_b\ll\Delta$, equation \ref{sconnn} can be simplified to:
\begin{equation} \Lambda_{n,k}=\frac{2r\Delta}{n+x/2}-2\sqrt{\frac{x}{2n+x}},\,\,\,\,\,\,\,\, n\neq0.
\end{equation}
The second term here reproduces the result for the first plateau, which is also the result of SCO in the absence of the magnetic field. The first term corresponds to the subleading correction to SCO in the presence of a magnetic field due to NNN hopping. We shall refer to the correction to SCO by $\delta\Lambda$:
\begin{equation}
    \delta\Lambda=\sum_{k}\sum_{n=1}^{N_{LL}}\frac{2r\Delta}{n+x/2}.
\end{equation}
Converting the sum over $n$ into an integral, one obtains:
\begin{equation}
    \delta\Lambda=2r\Delta\sum_k \ln\left(\frac{N_{LL}+x/2}{x/2}\right).
\end{equation}

Above, we have used the fact that in the quantized SCO regime, $1+x/2\simeq x/2$. First, we shall analyze the correction at vanishingly small magnetic fields. At such small magnetic fields we can get $N_{LL}$ using $N_{DS}/2$ ($1/2$ factor comes due to $N_{LL}$ being defined as number of LLs per valley) since the system essentially doesn't feel the effects of the magnetic field ($\phi_{tot}/\phi_0\ll1$). The sum over $k$ just gives 1 because the degeneracy isn't there at such small magnetic fields. Substituting the expression of $N_{DS}$, the expression boils down to:
\begin{equation}
    \delta\Lambda=2r\Delta \ln\left(1+N_{DS}/x\right).
\end{equation}

At very small magnetic fields corresponding to $\pi \phi_{tot}/\phi_0\ll\Delta^2/\alpha^2$, one can show using \ref{nds} that $N_{DS}/x\ll1$, and therefore one can approximate, $\ln(1+N_{DS}/x)\simeq N_{DS}/x$. It is important to note that this occurs at weak magnetic fields only if the sublattice potential is weak, i.e., $\Delta\ll\alpha$. When $\phi_{tot}/\phi_0\gg\Delta^2/\alpha^2$, i.e., $N_{DS}/x\gg1$, then $\delta\Lambda\sim r \Delta \ln1=0$. Consequently, the subleading correction due to NNN hopping exists only at weak sublattice potentials and vanishingly small magnetic fields. Substituting the expression for $N_{DS}$, we get:
\begin{equation}
    \frac{\delta\Lambda}{N_s}=
    \frac{1}{\sqrt 3}r\Delta\phi_b (\alpha/\Delta)^2.
\end{equation}

Therefore, the correction is approximately $\sim r\Delta|B|$, at vanishingly small magnetic fields corresponding to $\pi\phi_{tot}/\phi_0\ll\Delta^2/\alpha^2$ in the presence of weak sublattice potential, where $\Delta\ll\alpha$. This makes these corrections difficult to detect, even numerically. Above such small magnetic fields, the corrections disappear. Consequently, the effects of NNN hopping on SCO are very small and almost negligible.

\section{Benchmarks for Lattice Calculations}\label{benchmark}

\begin{figure}[H]
    \centering
    \includegraphics[scale=0.25]{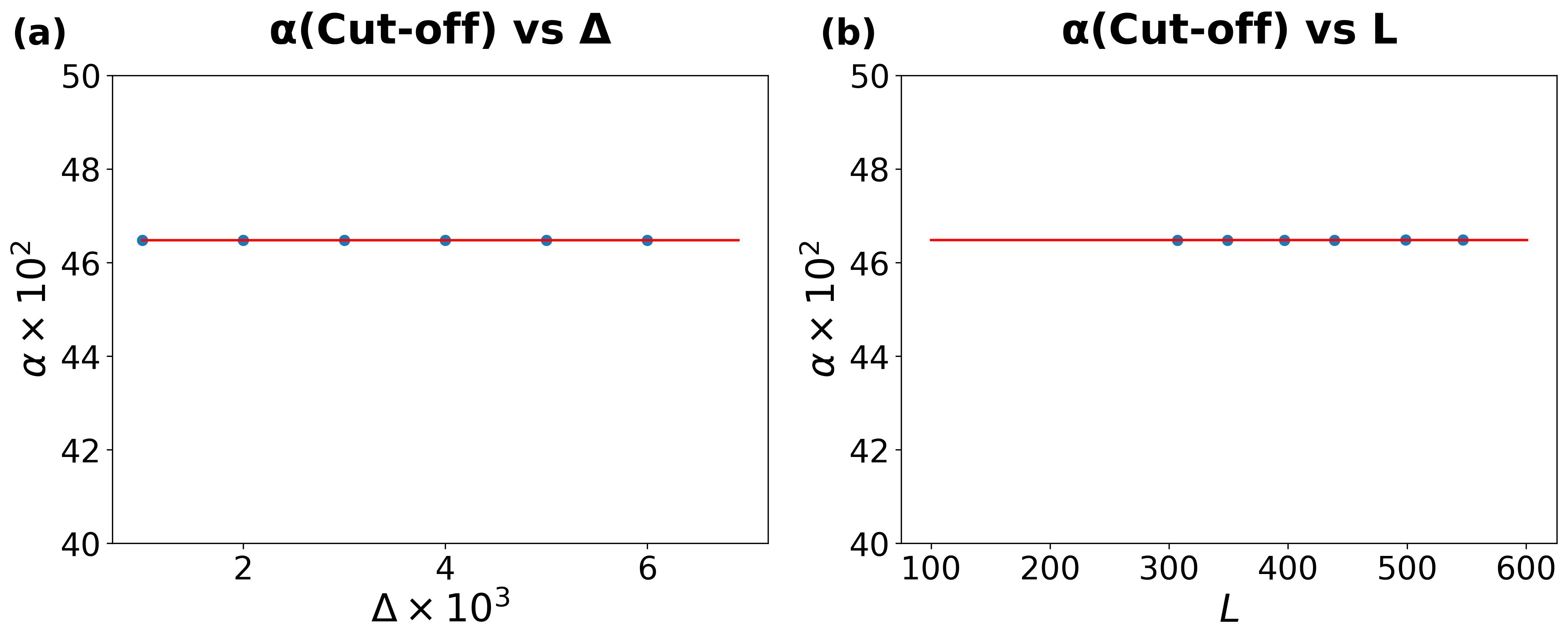}
    \caption{The left panel presents the dependence of the cutoff parameter $\alpha$ on the sublattice potential $\Delta$, while the right panel displays $\alpha$ as a function of the system size $L$. Our lattice calculations demonstrate that the value of $\alpha$ remains constant across different system sizes and sublattice potentials, confirming its robustness as expected for Dirac insulators.}
    \label{4}
\end{figure}

To validate the ultraviolet cutoff $\alpha \simeq 0.46$, determined from fitting the exact single-particle dispersion to the Dirac dispersion, we examine its behavior numerically for finite-size lattices across different system sizes and values of the sublattice potential $\Delta$, restricted to the regime $\Delta \ll \alpha$. Since the cutoff is an intrinsic property of the spectrum, it should remain independent of the system size. Likewise, in the weak-potential regime $\Delta \ll \alpha$, the cutoff is expected to be insensitive to variations in $\Delta$. The data presented in Fig.~\ref{4} corroborate these expectations.
\\

\bibliography{main}

\end{document}